\documentclass[iop]{emulateapj}
\usepackage{natbib}
\usepackage{mathptmx} 
\usepackage[T1]{fontenc}
\usepackage{graphicx}

\def\kms {{\mathrm{km}\,\mathrm{s}^{-1}}}
\def\caii {Ca\,\textsc{ii}}
\defcitealias{Zhang:2012}{ZSW12}
\defcitealias{Judge:2011}{JTL11}

\slugcomment{To appear in The Astrophysical Journal}

\usepackage[pdftitle={Quantifying Spicules},pdfauthor={Pereira, T. M. D., De Pontieu, B., \& Carlsson, M.}, pdfsubject={Solar spicules},
            pdfkeywords={Sun: atmosphere, Sun: chromosphere, Sun: transition region},
            colorlinks=true,urlcolor=blue,citecolor=blue,linkcolor=blue]{hyperref}
\shorttitle{Quantifying Spicules}
\shortauthors{Pereira, De Pontieu, \& Carlsson}

\begin{document}

\title{Quantifying Spicules}

\author{Tiago M. D. Pereira$^{1,2}$, Bart De Pontieu$^{2}$, Mats Carlsson$^{3,4}$}
\affil{$^{1}$NASA Ames Research Center, Mof{}fett Field, CA 94035, USA; \url{tiago.pereira@nasa.gov}\\
$^{2}$Lockheed Martin Solar \& Astrophysics Lab., Org. A021S, Bldg. 252, 3251 Hanover Street, Palo Alto, CA 94304, USA\\
$^{3}$Institute of Theoretical Astrophysics, P.O. Box 1029, Blindern, N-0315 Oslo, Norway\\
$^{4}$Center of Mathematics for Applications, University of Oslo, PO Box 1053, Blindern, 0316 Oslo, Norway
}

\begin{abstract}
Understanding the dynamic solar chromosphere is fundamental in solar physics. Spicules are an important feature of the chromosphere, connecting the photosphere to the corona, potentially mediating the transfer of energy and mass. 
The aim of this work is to study the properties of spicules over different regions of the sun. Our goal is to investigate if there is more than one type of spicules, and how spicules behave in the quiet sun, coronal holes, and active regions.
We make use of high-cadence and high-spatial resolution \caii\ H observations taken by \emph{Hinode}/SOT. Making use of a semi-automated detection algorithm, we self-consistently track and measure the properties of 519 spicules over different regions.
We find clear evidence of two types of spicules. Type I spicules show a rise and fall and have typical lifetimes of $150-400\;$s and maximum ascending velocities of $15-40\;\kms$, while type II spicules have shorter lifetimes of $50-150\;$s, faster velocities of $30-110\;\kms$, and are not seen to fall down, but rather fade at around their maximum length. Type II spicules are the most common, seen in quiet sun and coronal holes. Type I spicules are seen mostly in active regions. There are regional differences between quiet sun and coronal hole spicules, likely attributable to the different field configurations. 
The properties of type II spicules are consistent with published results of Rapid Blueshifted Events (RBEs), supporting the hypothesis that RBEs are their disk counterparts. For type I spicules we find the relations between their properties to be consistent with a magnetoacoustic shock wave driver, and with dynamic fibrils as their disk counterpart. The driver of type II spicules remains unclear from limb observations.
\end{abstract}


\section{Introduction}

Spicules can be seen almost everywhere at the solar limb. They are highly-dynamic, thin, jet-like features in the solar chromosphere. Spicules have been observed for a very long time and have been the subject of numerous reviews \citep[\emph{e.g.}][]{Beckers:1968,Beckers:1972,Suematsu:1995,Sterling:2000}. A fundamental part of the chromosphere, their origin and connection to the photosphere and corona have long been debated. 

Perhaps the most interesting aspect about spicules is their potential to mediate the transfer of energy and mass from the photosphere to the corona. This potential has been recognized early \citep{Beckers:1968, PneumanKopp:1978, Athay:1982}, but the lack of high-quality observations prevented a better understanding of spicules and their link to the corona \citep{Withbroe:1983}. The advent of the \emph{Hinode} Solar Optical Telescope \citep[SOT,][]{Kosugi:2007,Tsuneta:2008,Suematsu:2008} provided a quantum leap in the understanding of spicules and their properties. With its seeing-free high spatial and temporal resolution observations of the chromosphere, \emph{Hinode} showed much more dynamic spicules than previously thought and re-ignited the discussion on their potential to heat the corona \citep{DePontieu:2007, DePontieu:2009, DePontieu:2011}. %

Crucial for the understanding of the origin and effects of spicules is a characterization of their properties. While there are several recent studies of the properties of spicules using \emph{Hinode}/SOT \citep{DePontieu:2007,Anan:2010,Zhang:2012}, they are either brief first results papers and/or cover only one or two data sets. A comprehensive study of the properties of spicules in different solar regions is missing, and that is the aim of this work. Our goal was to obtain statistically-significant measurements of the properties of spicules in three key regions: quiet sun, coronal holes and active regions. In this we make use of the wealth of high-quality data taken with \emph{Hinode}/SOT, and of automated detection algorithms.

The outline of this paper is as follows. A description of the observations is made in Section~\ref{sec:obs}, followed by a description of the data analysis and automated spicule detection in Section~\ref{sec:spic_det}. Our results are presented in Section~\ref{sec:results}, and in Section~\ref{sec:discussion} we discuss their consequences, including the types of spicules found and in which regions, and a comparison with other results from the literature. Finally, our conclusions are summarized in Section~\ref{sec:conclusions}.

\section{Observations}
\label{sec:obs}

\subsection{Data Sets}

We make use of a series of data sets obtained with the \emph{Hinode} SOT Broadband Filter Imager \citep{Kosugi:2007,Tsuneta:2008,Suematsu:2008}, in the Ca~\textsc{ii} H (396.85 nm) filter. The data sets are from limb observations, as on the disk the Ca~\textsc{ii} H filter radiation comes mostly from the photosphere \citep{Carlsson:2007} and spicules are hardly visible. The selected data sets are summarized in Table~\ref{tab:obs}. To capture in detail all the stages of spicules, we selected data sets with the highest cadence available: in most cases 4.8 seconds or less. The pixel size of the images used was $0.\arcsec109$; some data sets had a pixel size of $0.\arcsec054$ and were binned to increase the signal to noise. The image sizes were between 512 to 1024 pixels in either dimension. 

Three types of regions were selected: quiet sun, coronal hole, and active regions. Considerable care went into identifying which data sets belonged to these regions. Active regions were identified with the help of \emph{Hinode}/XRT data, where they stand out in X-ray emission. To avoid line-of-sight confusion and provide an accurate determination, we also rotated the limb positions back to disk positions of the active region a few days earlier and inspected the disk data. Some seemingly quiet patches around those regions were also classified as active because it was not possible to rule out light contamination from active regions (important in these optically-thin diagnostics at the limb). Because of the proximity of active regions, quiet sun data sets with the required cadence and resolution were particularly hard to find. In many \emph{Hinode} observations around the equatorial limb there is either an active region visible or one is about to emerge. For example, \citet[hereafter \citetalias{Zhang:2012}]{Zhang:2012} classify the \mbox{2006-11-21T22:57} dataset as quiet sun, while we find it to clearly be an active region dataset (\emph{e.g.} presence of active region plage in the foreground, and very bright XRT images). Hence, only sets found to be unambiguously quiet were classified as quiet sun.

\begin{deluxetable}{lrrrr}
\tablecaption{Observational data sets.\label{tab:obs}}
\tablehead{
\colhead{Starting time} & \colhead{x coord.} & \colhead{y coord.} & \colhead{Cadence} & \colhead{Duration} \\
     & \colhead{(arcsec)} & \colhead{(arcsec)} & \colhead{(s)} & \colhead{(min)}}
\startdata
Quiet Sun & & & & \\
2007-04-01T14:09 & $ 900$ & $-288$ & $4.8$ & $55$ \\
2007-04-01T15:09 & $ 933$ & $-145$ & $4.8$ & $55$ \\
2007-08-21T13:35 & $ 945$ & $   0$ & $4.8$ & $204$ \\
2011-09-16T03:00 & $-955$ & $ 259$ & $3.2$ & $66$ \\[0.1cm]
\tableline
Coronal Hole & & & & \\
2007-02-09T11:43 & $980$ & $-40$  & $8.0$ & $382$ \\
2007-03-19T15:30 & $0$   & $-968$ & $4.8$ & $60$ \\
2011-01-29T11:06 & $268$ & $-945$ & $1.6$ & $53$\\[0.1cm]
\tableline
Active Region & & & & \\
2006-11-21T22:57 & $960$ & $-90$ &   $8.0$ & $202$ \\
2007-02-16T13:10 & $939$ & $-193$ & $11.2$ &  $89$ \\
2007-05-25T06:33\tablenotemark{a} & $925$ &  $25$  &   $4.8$ & $16$ \\ 
2009-07-11T22:51 & $847$ & $-434$ &  $3.2$ & $30$
\enddata
\tablenotetext{a}{This set was only used for the faint spicules discussed in section~\ref{sec:ar_spic}.}
\end{deluxetable}
\vspace{0.1cm}

\subsection{Data Reduction}

The original data were reduced using standard methods such as the IDL routine \texttt{fg\_prep.pro} for dark current subtraction and flat fielding (available in solarsoft, \url{http://www.lmsal.com/solarsoft}). A cross-correlation technique \citep[as in][]{DePontieu:2007} allowed for a precise co-alignment of the images. To enhance the visibility of spicules we apply a radial density filter \citep{DePontieu:2007,Okamoto:2007}. Additionally, we convolve the images with a diagonal difference kernel (emboss) and add it back to the radial-filtered version to enhance the edges of spicules and help with the automated detection.

\section{Data Analysis and Spicule Detection}
\label{sec:spic_det}

\subsection{Semi-automated Procedure}
\label{sec:detection}

\begin{figure}
\begin{center}
\includegraphics[width=0.465\textwidth]{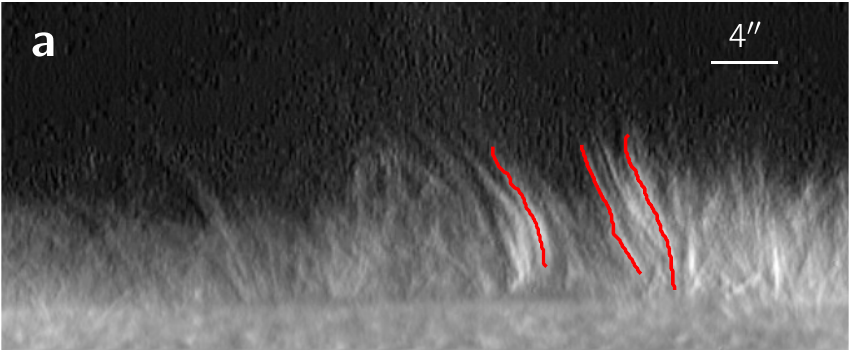}\\
\includegraphics[width=0.465\textwidth]{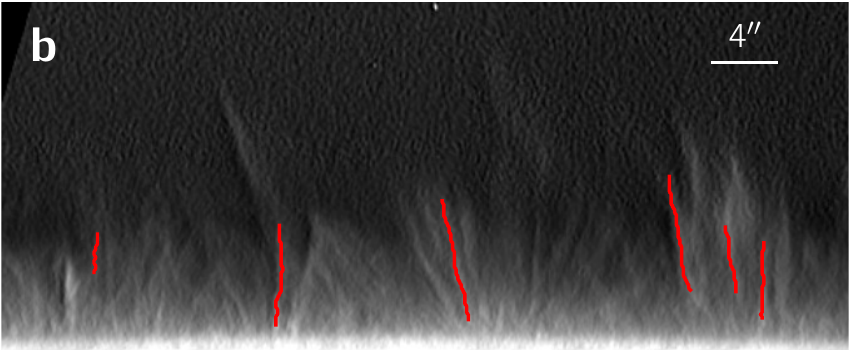}\\
\includegraphics[width=0.465\textwidth]{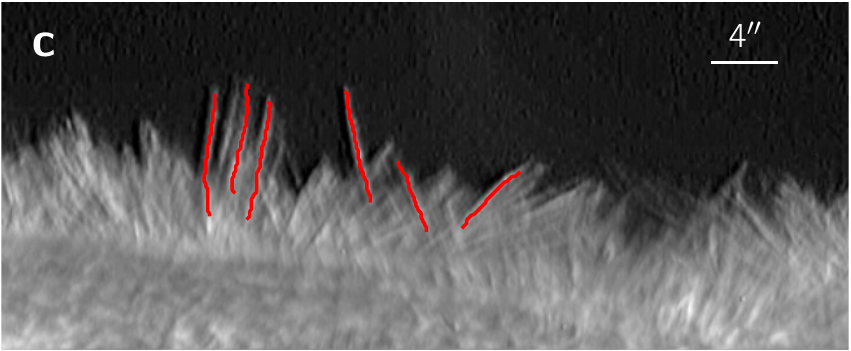}
\end{center}
\caption{Sample spicule images and automated detection for: (a) quiet sun data from 2007-08-21, 14:03:33 UT; (b) coronal hole data from 2011-01-29, 11:12:12 UT; (c) active region data from 2006-11-21, 23:08:51 UT. Each panel is approximately $53\arcsec\times22\arcsec$, and the approximate solar (x, y) coordinates for the lower left corners are ($936.8\arcsec$, $-26.4\arcsec$), ($256.1\arcsec$, $-951.3\arcsec$), ($944.4\arcsec$, $62.1\arcsec$), respectively for (a), (b), and (c). Radial density and emboss filters were applied to the images, and the intensity scaling (also as a function of height) differs for each image. Red lines denote spicules detected automatically in these images, which provided the starting point for our measurements.\label{fig:sd}}
\end{figure}

To achieve our goal of measuring the properties of spicules, an automated procedure was necessary. It saved a lot of manual work that would be required otherwise, but most importantly it allowed us to reliably track the significant transverse motions of spicules \citep[caused by Alfv\'enic waves,][]{DePontieu:2007} and keep the spicule detection biases more limited and constant between data sets. The transverse motion tracking enables more accurate measurements of spicules than an x-t diagram analysis \citep[\emph{e.g.}, by][]{DePontieu:2007} \footnote{Note that \citet{DePontieu:2007} used the x-t diagram analysis only on spicules that did not show significant transverse motions, thus minimizing the errors on spicule parameters, but also limiting the sample size of the number of spicules}. Our analysis can be divided in two steps: spicule detection and extraction of its properties. 

The spicule detection consists of finding the temporal evolution of the position of spicules using the time series of edge-enhanced images. The automated detection algorithm used is based on the method used by \citet{vdVoort:2009} for RBEs. Different spicules are detected in each image, then linked across several images (until no longer detected), and a chain of events is produced for each spicule. 

To limit the number of false positives (usually caused by line-of-sight superposition) in the spicule detection, we adopted a few limits on the parameters for the detection algorithm. A minimum value for the maximum length of spicules was set at $\approx 1.185\;$Mm (15 pixels). Spicules whose closest point to the limb was more than $\approx 6.3\;$Mm (80 pixels) from the limb were not considered, along with spicules with an inclination of more than 75$^{\mathrm{o}}$. For the linkage in time of single events, we rejected events that had more than 30\% of their points further than $\approx 0.3\;$Mm (4 pixels) from the event detected in the previous time step. This imposes a limit on the maximum transverse speed we can detect, which depending on the cadence will go from $\approx 28-198\;\kms$ (but for the typical cadence of 4.8~s it is $\approx 66\;\kms$). Nevertheless, these values are lower bounds on the true limit, because the transverse velocity is actually measured at a height in the middle of the spicule by averaging several points that {\em can} be further than 4 pixels apart.

\begin{figure*}
\includegraphics{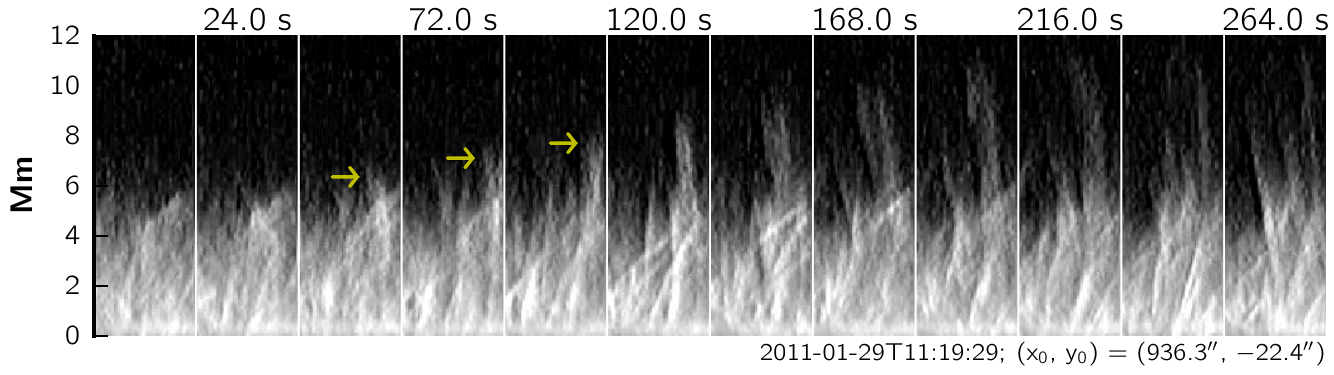}\includegraphics{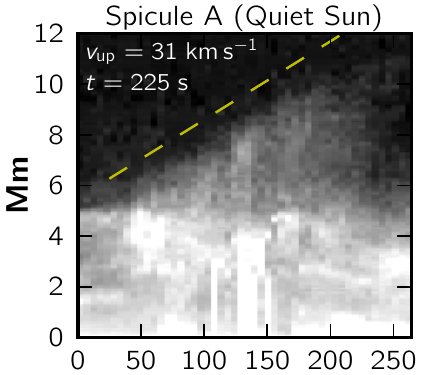}\\
\includegraphics{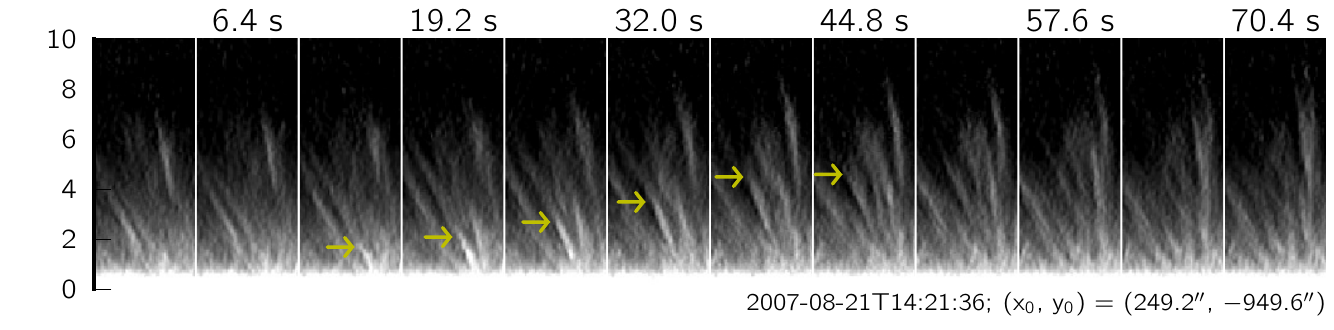}\includegraphics{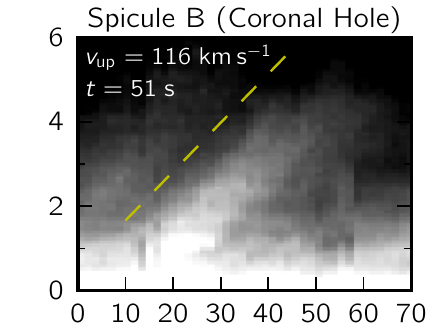}\\
\includegraphics{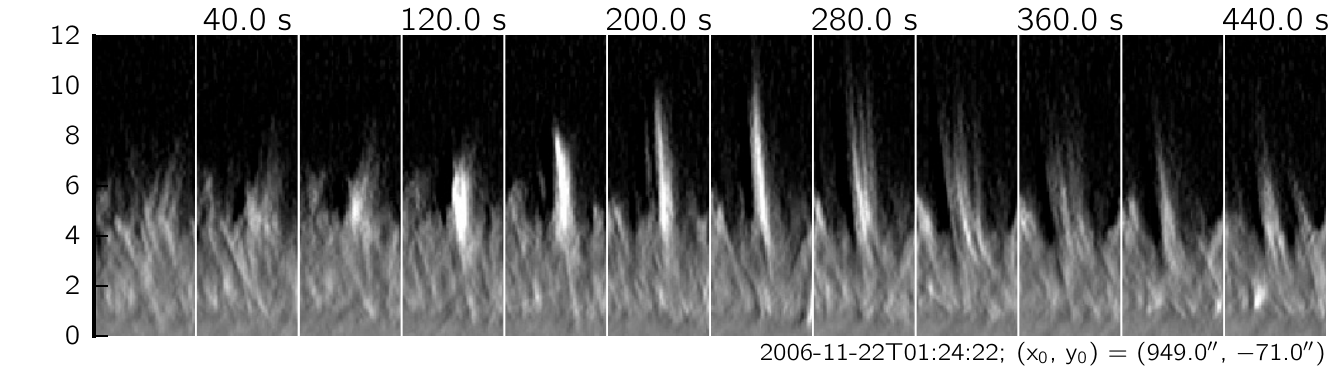}\includegraphics{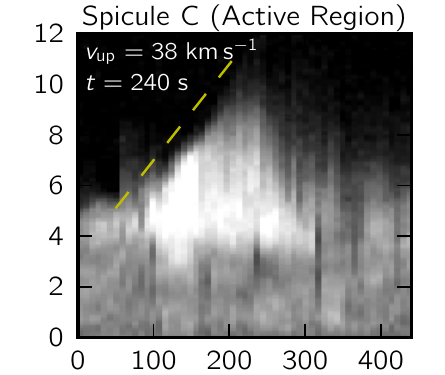}\\
\includegraphics{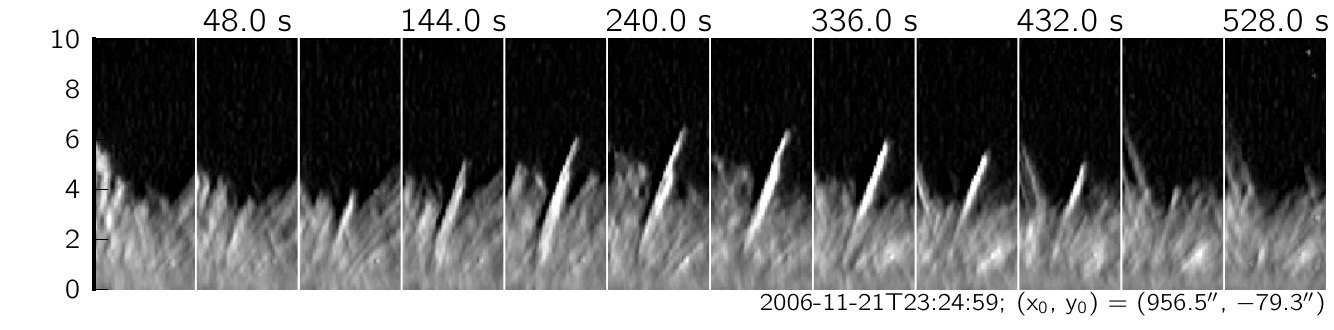}\includegraphics{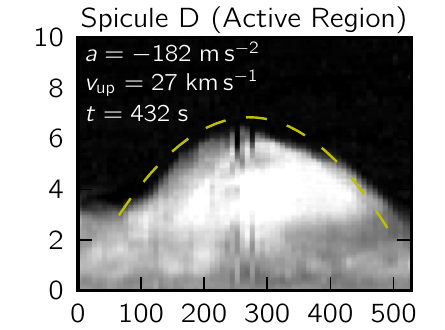}\\
\includegraphics{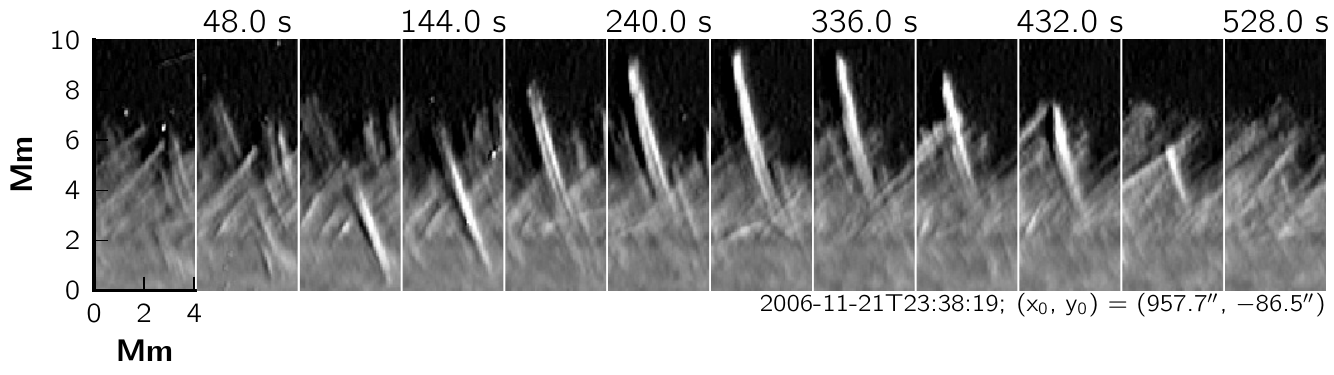}\includegraphics{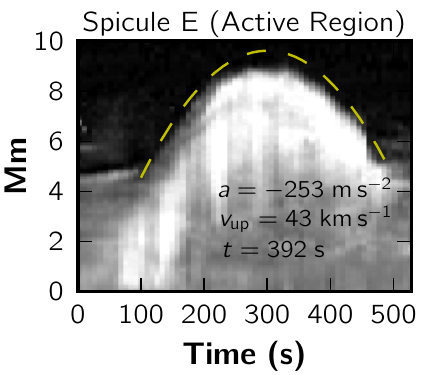}
\caption{Diagrams of spicule evolution. For each row, the left column shows a time sequence of Ca\,\textsc{ii} filtergrams in a region around a spicule. The right column shows the corresponding space-time diagram, showing the intensity along the spicule for each time step. Arrows are shown in some panels to clarify the position of the spicule being studied. The space-time diagrams are built by tracking the transverse motions of the spicules. The linear or parabolic fits are shown in a dashed line (lines have been shifted up for clarity). Some of the measured parameters for each spicule are listed. The height scales are not directly comparable between the two columns: on the left it represents distance in the image (with an arbitrary zero point); on the right it represents the length along the spicule, measured from its (approximate) starting point. Spicule A was observed on quiet sun data, spicule B was observed on coronal hole data, and spicules C, D, and E were observed on active region data. The observing time of the first image and the approximate solar (x,y) coordinates of its lower left corner are shown below each sequence of images. \label{fig:spic_evo}}
\end{figure*}

Because no algorithm is perfect at detecting spicules, human intervention is necessary at several stages. The first of these is to manually discard false chains of events (\emph{e.g.} spicules that are incorrectly linked, or artifacts that were linked but are not spicules). In Fig.~\ref{fig:sd} example images are shown with the detected spicules (after human intervention). When necessary, the images shown in Figs.~\ref{fig:sd} and~\ref{fig:spic_evo} have been rotated to show the solar limb at the bottom and the spicules reaching up.

With the spicule detection we obtain the coordinates of several spicules across a series of images. The next step is to measure the properties of each spicule. Ten quantities were extracted: lifetime, maximum velocity, maximum length, maximum height, width, inclination, transverse displacement, transverse speed, scale height, and deceleration (for spicules that were observed to rise and fall). How these were calculated is detailed below.

\subsection{Extracted Quantities}

\paragraph{Lifetime}
The automatic detection provides an estimate of when the spicule is born and when it disappears. However, this is very approximate in some cases (\emph{e.g.} when the spicule is smaller than the imposed minimum length for detection, or when the detection struggles due to a nearby spicule crossing over or overlapping). By visual inspection, a time search is done around the images where the spicule is automatically detected. Here, we identify the images where the spicule is first and last seen. The extracted lifetime is the time interval between these two images.

\paragraph{Maximum Length and Height}
The maximum length of a spicule was extracted manually. By visual inspection of the life of the spicule, the image where it reaches its maximum extent is found, and the maximum length is measured along the spicule from the top of the spicule until the limb. Here it is assumed that the spicules always connect to the limb, although their starting point is not always evident. For the spicules whose `main body' does not connect to the limb, the length of the distance to the limb was approximated. The maximum height is measured similarly, but instead of along the spicule it is the shortest distance from the spicule's highest point to the limb. 

\paragraph{Inclination} 
The inclination of a spicule is measured by doing a linear fit to the detected spicule points and calculating the angle with the normal to the limb (\emph{i.e.}, the radial direction). For the inclination statistics we include all the frames in which the spicule was detected. The inclination is thus an average inclination over the lifetime of the spicule.

\paragraph{Maximum Velocity and Deceleration}
The maximum velocity of the spicules is extracted with the help of space-time diagrams. These diagrams are built by obtaining the image intensity along the length of the spicule, for each image where the spicule is detected. For a given image, the position of the spicule is obtained from the automatic detection, which has an inherent uncertainty. To minimize the statistical fluctuations, a linear fit (in space) is done to the spicule coordinates. The image intensity is then interpolated for the pixel positions derived from the linear fit, as well as the two neighboring positions obtained by shifting the points by $\pm1$ pixel in the horizontal direction (to maximize the signal, the interpolated intensity is averaged for these three positions).

By using a linear fit one is assuming that spicules are not curved. This is not always the case, and the method struggles for curved spicules. Nevertheless, it works robustly for most spicules. It was found that extending the parameter space to define the position of the spicules (\emph{e.g.} by fitting a higher-order polynomial) brings additional problems such as ill-defined distance from the spicule to the limb. Hence, the linear fit was used, with the caveat that strongly curved spicules could not be reliably analyzed. 

Interpolating the original image yields the intensity values along the length of the spicule, for each time step. As spicules also move with a velocity transverse to their own axis, our method tracks the spicule for each time step, and thus takes the transverse motion into account. Using the intensity measurements along the spicule a space-time diagram is built, showing the brightness of the spicule during its life. The trajectory of a spicule in the x-t space is one of its most basic properties. Given that in this space most spicules seem to either describe a parabola (spicules that rise and fall) or a line (spicules that rise and fade), we will henceforth use the terms `parabola' or `linear' spicule to distinguish between the two behaviors. Examples of these x-t diagrams are shown in the right column of Fig.~\ref{fig:spic_evo}, for spicules with parabolic and linear trajectories in the x-t space. With this diagram, the highest point of the spicule is manually identified for each time step. A fit is made to these x-t points, from which the spicule velocity is extracted. For the spicules with a parabolic trajectory a parabola is fitted (from which the deceleration is also derived). For the other spicules, a line is fitted. In some cases (\emph{e.g.} because of superposition) the highest point of the spicule is not discernible for several time steps. In these cases, a point at a different height of the spicule is tracked instead. 

\paragraph{Width}
The width or diameter of a spicule is defined as the FWHM of its intensity profile in a direction perpendicular to the spicule axis. We measure one value of width per spicule, at the instant when it reaches its maximum length. The intensity profile across the spicule is evaluated at a point corresponding to 75\% of its height. Measuring the width at this height is a tradeoff: in the lower part of the spicule the field is more crowded with spicules (affecting the background intensity), and at the top the spicule is ill-defined; at 75\% of the height one has a better chance to derive the spicule over a darker background. To obtain the intensity profile we interpolate the image in the direction normal to the spicule, and average it at three different heights (at 75\% height $\pm0.5$ pixels) to increase the S/N. To extract the FWHM from the intensity profile we fit a Gaussian function. For each spicule the fitting range is manually defined, and a good fit is visually confirmed. With this procedure we make sure to only fit over the spicule component we are interested in, when a spicule has nearby companions or has split in a few components. Some spicules ($\approx 10\%$) had irregular intensity profiles from which a reliable width could not be extracted.

\paragraph{Transverse Speed and Displacement}
The transverse speed and displacement are measured from a constant height, typically in the central part of the spicule. A central height point of a spicule is found by averaging the detected spicule height points in time. The spicule displacement at this height is measured for each image and the transverse speed calculated. The transverse displacement is measured as the maximum transverse motion that the spicule has suffered in this central point over its life time (\emph{i.e.}, from the two most extreme positions).

\paragraph{Scale Height}
The scale height is defined as the distance over which the filtergram intensity decreases by a factor of $e$. Using the original flat-fielded images (without the radial and emboss filters), we calculate two types of scale heights: a `global' scale height averaged in angular distance over the whole image; and a mean scale height for each spicule. 

Assuming an exponential drop-off of the intensity at the solar limb, the global scale height $H_s$ is a function of the height or radial distance $r$, and is calculated using the expression:
\begin{equation}
\label{eq:scaleh}
H_s(r) = -\left(\frac{\mathrm{d}}{\mathrm{d}r}\ln I(r)\right)^{-1}.
\end{equation}
To calculate the scale height for each observational set, we start by applying a geometric transformation to the images so that each column represents the radial intensity profile at a direction perpendicular to the limb (i.e., we ``straighten out the limb''). Then, the radial intensity profile is averaged over the angular direction and over time windows of approximately one minute. The position of the limb is calibrated from the inflection point in this mean radial profile, and the global scale height calculated from Eq.~(\ref{eq:scaleh}).

A mean scale height for each spicule is calculated in the following manner. First, the intensity profile $I(d)$ along the spicule is obtained by interpolating the image in the coordinates of the spicule at the point of its maximum length. As done for the space-time diagrams, the intensity is averaged over the original points plus the two neighboring pixels. We then fit a line to $\ln{}I(d)$, from which the mean scale height is $\overline{H}_s \approx -1/m$, where $m$ is the line fit slope. This fit is done for $d < 6$\,Mm, because after that point the signal starts to be dominated by noise. Finally, the scale height along the radial direction is obtained by multiplying the fit slope by the cosine of the angle between the spicule and the normal to the limb.

\subsection{Selection Effects and Errors\label{sec:selerr}}

Detecting and measuring spicules in Ca\,\textsc{ii} H images is a difficult task. Not only are spicules faint against the background, they are ubiquitous at the solar limb with a large degree of superposition. Superimposed spicules move in front or behind other spicules, often with similar trajectories and properties, sometimes with oblique angles, and often with seemingly similar intensities that almost seem to blend together. They are difficult to disentangle with the human eye, even more for the detection algorithms. Adding the fact that many spicules are short-lived and with significant transverse motions, their detection is truly challenging. In particular for data sets with longer cadences where the transverse motions are not so easily traceable, and when a spicule moves or vanishes it can be confused with another spicule that has moved to its position in the next frame. One should keep in mind all these detection difficulties when contemplating the results. They introduce a moderate degree of subjectivity in the extraction of the spicule properties, and are subject to different selection effects and errors. These are discussed below.

Superposition of many spicules causes the majority of selection effects. To be detected by the automated procedure, a spicule needs to stand out from the background. When a spicule is observed against a background of many overlapping or closely spaced spicules, its contrast from the background is reduced and therefore it is harder to detect. Thus, brighter, thicker, and longer spicules that rise above the `forest of spicules' will be detected more often, leading to a selection effect. This is particularly true in active regions, where the `forest of spicules' is denser, and less of a problem in coronal holes. Superposition of oblique spicules, even if the field is not too crowded, can also result in detection difficulties. The detection algorithm assumes mostly linear shapes for spicules. When two spicules of similar intensities overlap, the detection algorithm will have trouble disentangling them and in some cases can try and join two oblique spicule segments, which will then be rejected because of its non-linear shape.

Foreground spicules are easier to detect. However, spicules closer to the solar disk will seldom be detected -- and in many cases imperceptible. This is because photospheric light increasingly dominates the Ca\,\textsc{ii} H bandpass filter on the solar disk. This difficulty in seeing spicules on the disk influences the measurements of spicules seen at the limb. Spicules whose footpoints are further from the limb (closer to the observer and inside the solar disk), are prone to have their length measurement underestimated. Similarly, the measurement of lifetime will also be underestimated. In a few parts of some datasets the field-of-view of SOT is not enough to see the top of the tallest spicules, or is partially cut because the solar limb is not aligned with the image axes. This prevents one from measuring the heights of the tallest spicules. For the \mbox{2007-03-19} dataset the maximum height imposed by this restriction was about $11.5$~Mm; most of the other datasets have a maximum limit of $14$~Mm or more. The vast majority of the observed spicules have smaller heights than these values, so the effects of this restriction are limited.

\begin{deluxetable}{lrr} 
\tablecaption{Detected spicules\label{tab:spic}.}
\tablewidth{0pt}
\tablehead{
\colhead{Dataset} & \colhead{Number of Spicules} & \colhead{Linear Fraction}\\
       & \colhead{(Linear/Parabola)}  & \colhead{(\%)}  }
\startdata
Quiet Sun &  $\mathbf{174/3}$ & $\mathbf{98}$ \\
2007-04-01T14:09 & $24/0$  & $100$ \\
2007-04-01T15:09 & $47/1$  & $98$ \\
2007-08-21T13:35 & $62/2$  & $97$ \\
2011-09-16T03:00 & $41/0$  & $100$ \\
\tableline
Coronal Hole & $\mathbf{170/2}$ & $\mathbf{99}$ \\
2007-02-09T11:43 & $59/2$  & $97$  \\
2007-03-19T15:30 & $60/0$  & $100$ \\
2011-01-29T11:06 & $51/0$  & $100$ \\
\tableline
Active Region & $\mathbf{58/112}$ & $\mathbf{34}$ \\
2006-11-21T22:57 & $19/79$  &  $19$ \\
2007-02-16T13:10 & $39/18$  &  $68$ \\
2009-07-11T22:51 & $0/15$   &  $0$ \\
\tableline
\textbf{Total} & $\mathbf{402/117}$ & $\mathbf{77}$
\enddata
\end{deluxetable}

The top end of spicules is ill-defined. In most cases there is no sharp reduction in intensity but a gradual falloff. The exact point where a spicule ends, important for length and velocity measurements, has thus a degree of subjectivity that depends on the person making the estimate, on the background level, the background intensity, and signal to noise ratio of a given image. Care was taken in homogenizing these factors to keep the bias constant among data sets. Nevertheless, it leads to uncertainties in both velocity and length. The velocity estimation, as derived from the x-t diagrams, also suffers from the overlapping of spicules -- it may affect the visibility of its top end. The last important source of uncertainty in the velocity is the assumption of linear shape when calculating the \mbox{x-t segments}.

The cadence of the observations also adds uncertainty to the spicule measurements. It is more relevant in the determination of the lifetime (where a spicule may disappear between frames when observed with long cadences), and of the transverse velocity. In the latter case, because there is no information about the transverse velocity in between images, it is possible to miss the detection of rapid bursts in transverse motion if they were to happen on a very short time-scale.

When the detection algorithm connects spicules across time (\emph{i.e}, in consecutive images), the number of false positives is much larger for spicules that are detected only in a few images. This happens because false positives tend to happen randomly in a given image, and the likelihood of one being found in the same area in the next few images decreases sharply. To filter out this high number of false positives, spicules that are detected in less than four images are not considered. This introduces a selection effect, in which shorter-lived spicules are rejected. Given that the longest cadence used was \mbox{11.2\,s} and that the cadence for most data sets was \mbox{$8\;$s} or less, spicules that last for less than \mbox{$32-45\;$s} were rejected. However, comparing spicule lifetimes from coronal hole datasets with different cadences (from 1.6 to 8~s) one obtains very similar distributions, suggesting that the impact of this selection effect is small.
Short spicules that are rejected on the basis of length (see Section \ref{sec:detection}) often have shorter lifetimes and are missing from our distributions. Proper inclusion could skew the measured lifetimes towards lower values.

Another problem affecting the temporal linkage of detected spicules is what happens when a spicule is not detected in one or more intermediate images, but it is detected in subsequent images. This could introduce a selection effect, where the longer lived spicules would have their lifetime underestimated. The detection algorithm tries to correct for this by searching for the spicule in the subsequent $t+1$ image, when detection fails at an image take in instant $t$. It could also be made to search in the $t+2$, \ldots, $t+i$ images, however this was found to introduce more false detections (other spicules that move into a similar position and are incorrectly linked). In any case, each spicule is manually followed in images before and after it is detected automatically and the lifetime is considered by visual inspection, greatly minimizing this selection effect.

\section{Results}
\label{sec:results}

\subsection{Spicule Properties}

We present results for the three regions considered: quiet sun, coronal holes, and active regions. The data sets employed and number of spicules detected are summarized in Table~\ref{tab:spic}. Overall, more than 500 spicules were tracked and measured. We include only spicules whose life history could be traced across several frames. Spicules detected in only one frame, partial detections, and spicules whose ascending phase is not seen were not considered. This helped maintain a high quality sample of spicules, at the cost of not detecting many of the spicules visible in filtergrams (mostly because of the limitations in automated detection). This cost was offset and the statistics improved by using several data sets, and more than 100 spicules for each region.

\begin{figure*}
\begin{center}
\includegraphics[scale=0.95]{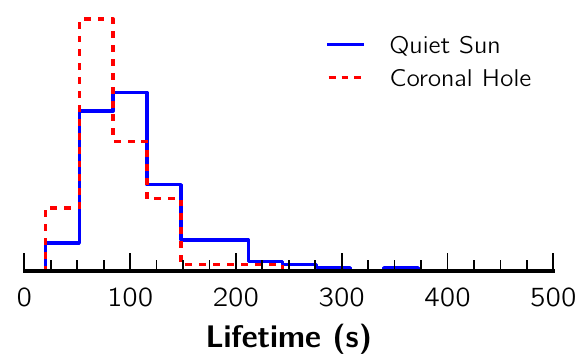}\includegraphics[scale=0.95]{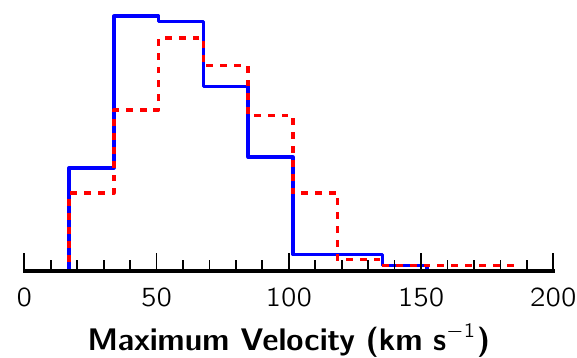}\includegraphics[scale=0.95]{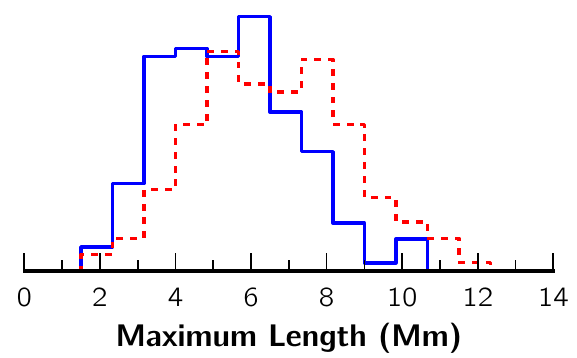}\\
\includegraphics[scale=0.95]{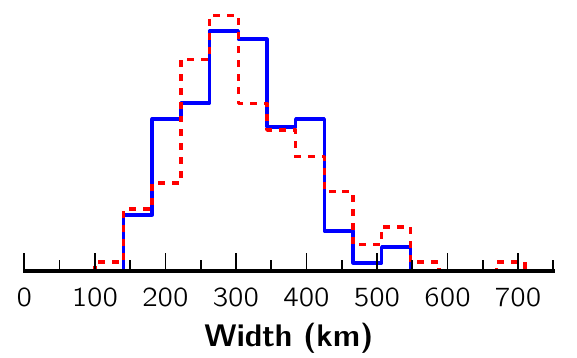}\includegraphics[scale=0.95]{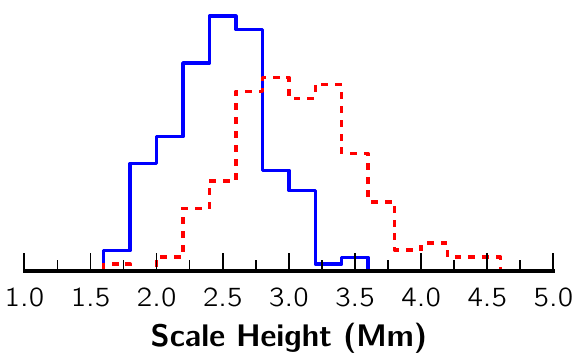}\includegraphics[scale=0.95]{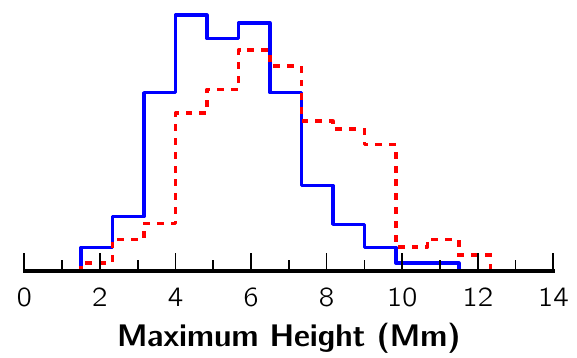}\\
\includegraphics[scale=0.95]{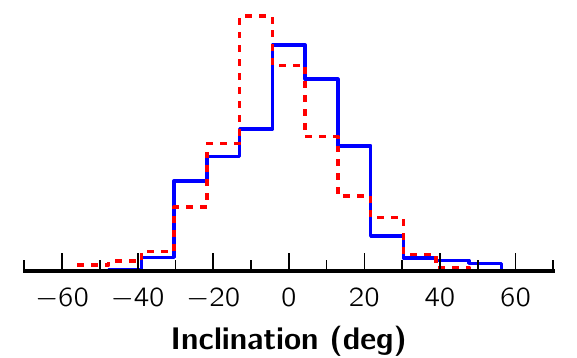}\includegraphics[scale=0.95]{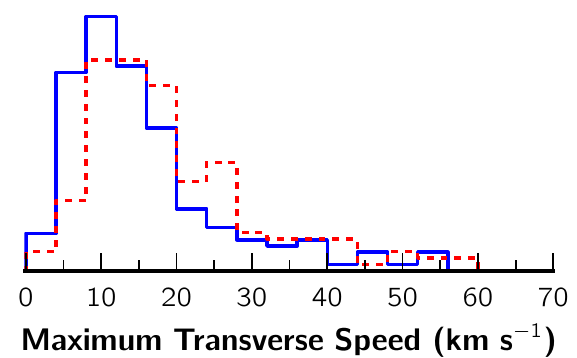}\includegraphics[scale=0.95]{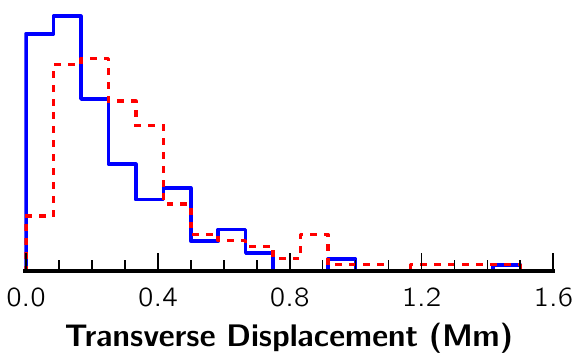}
\end{center}
\caption{Normalized distributions of spicule properties in the quiet sun (blue solid) and coronal hole (red dashed).\label{fig:stat_qsch}}
\end{figure*}

\begin{figure*}
\begin{center}
\includegraphics[scale=0.95]{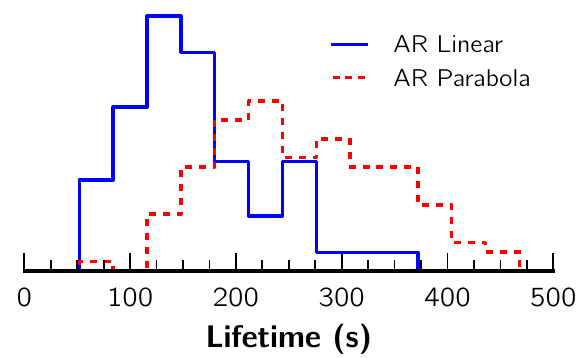}\includegraphics[scale=0.95]{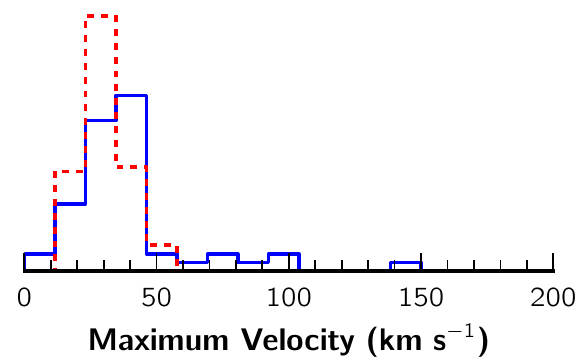}\includegraphics[scale=0.95]{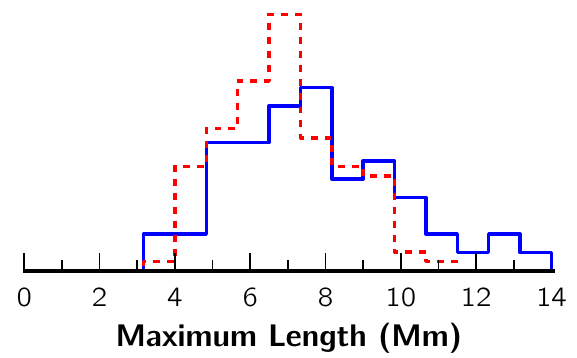}\\
\includegraphics[scale=0.95]{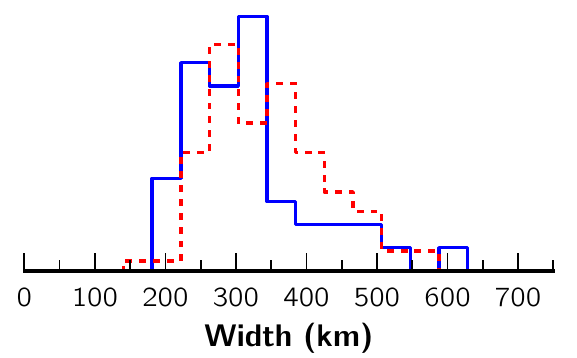}\includegraphics[scale=0.95]{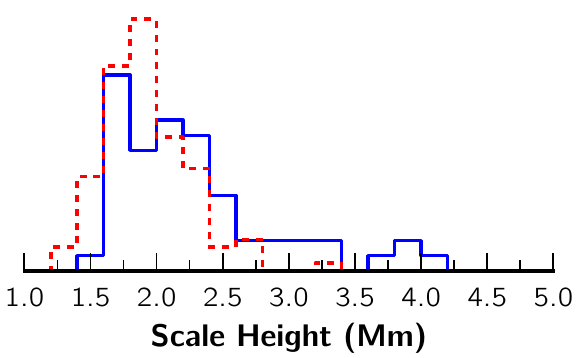}\includegraphics[scale=0.95]{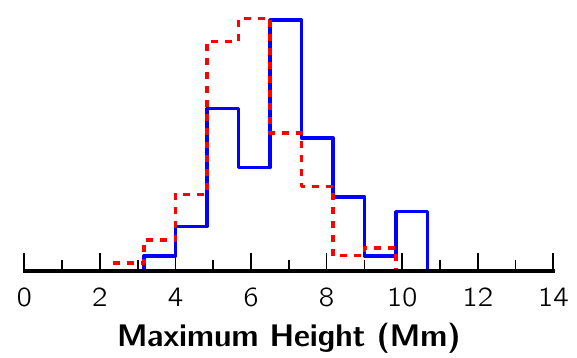}\\
\includegraphics[scale=0.95]{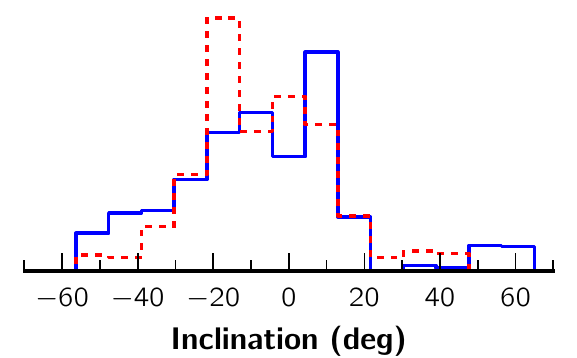}\includegraphics[scale=0.95]{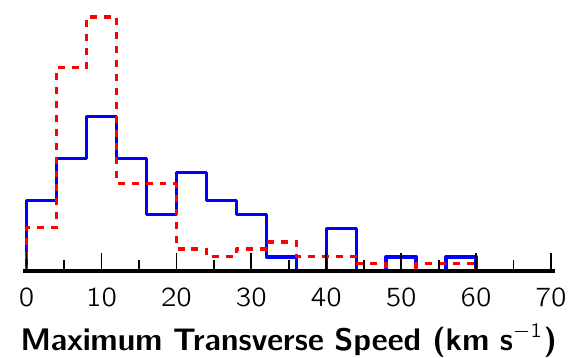}\includegraphics[scale=0.95]{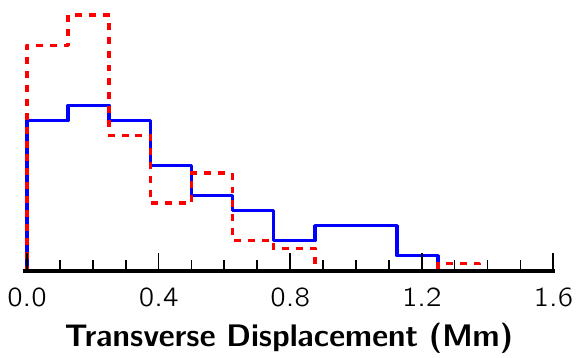}
\end{center}
\caption{Normalized distributions of spicule properties in active regions, for parabola spicules (red dashed) and linear spicules (blue solid).\label{fig:stat_ar}}
\end{figure*}

The most basic distinction in the spicules observed in the \mbox{Ca\,\textsc{ii} H} filtergrams concerns their motion. Some spicules show only upward movement (away from the solar surface) while others move up and down in a parabolic trajectory in the x-t space. Based on their trajectory in the x-t space, we have classified spicules as `linear' or `parabola'. 
The linear spicules are dominant in most of the sun except active regions. Conversely, parabola spicules are seldom found outside of active regions. For active regions, we separated the results of these two populations of spicules. The few parabola spicules found in quiet sun and coronal holes have properties very similar to those found in active regions (lifetimes $> 200\;$s; velocities $<50\;\kms$). Given their scarcity, only linear spicules were included in the statistics for quiet sun and coronal hole spicules. 

\begin{figure*}
\begin{center}
\includegraphics[scale=0.95]{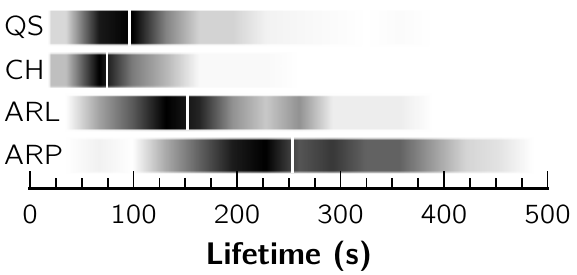}\includegraphics[scale=0.95]{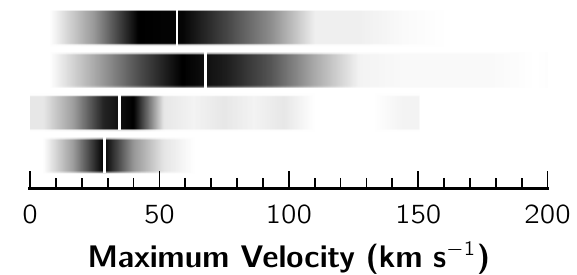}\includegraphics[scale=0.95]{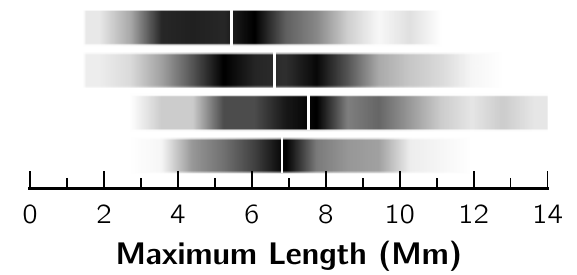}\\
\includegraphics[scale=0.95]{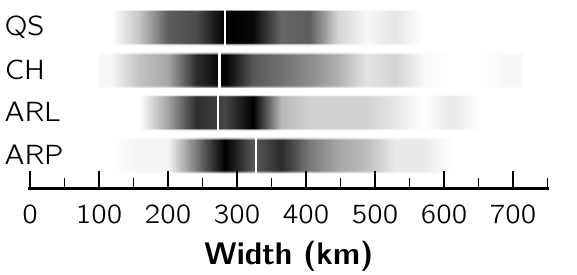}\includegraphics[scale=0.95]{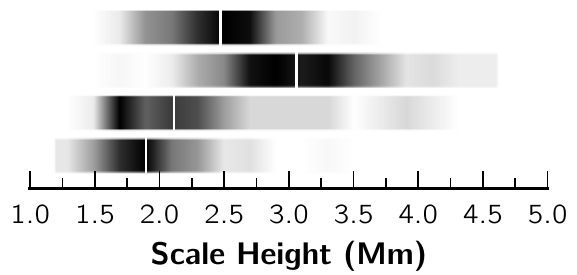}\includegraphics[scale=0.95]{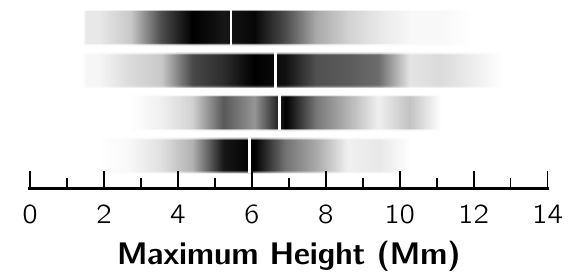}\\
\includegraphics[scale=0.95]{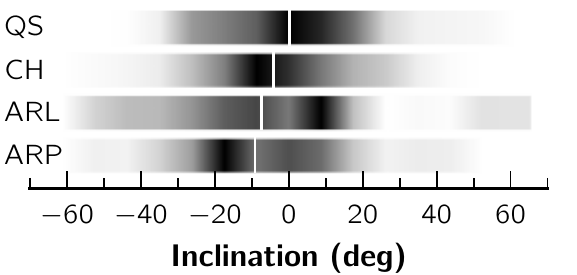}\includegraphics[scale=0.95]{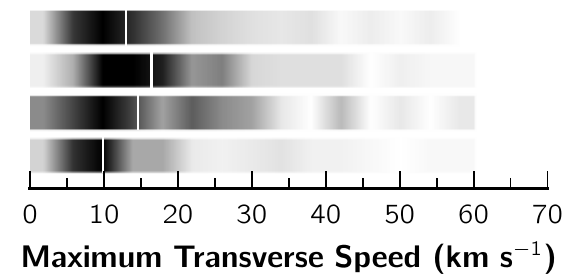}\includegraphics[scale=0.95]{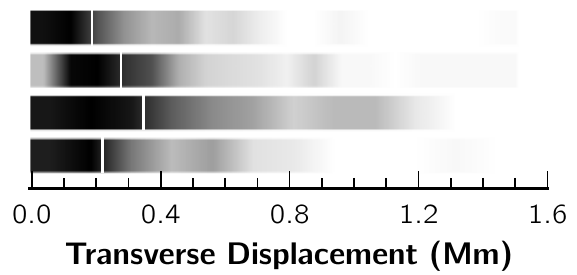}
\end{center}
\caption{Density diagrams for spicule properties. Luminance interpolated from histograms in Figures~\ref{fig:stat_qsch} and \ref{fig:stat_ar}; darker means more frequent. For each quantity diagrams correspond to, from top to bottom: quiet sun (QS), coronal hole (CH), active region linear (ARL), active region parabola (ARP). White vertical lines denote medians.\label{fig:spic_density}}
\end{figure*}

Histograms for nine key quantities (lifetime, maximum velocity, maximum length, maximum height, inclination, maximum transverse speed, transverse displacement, and scale height) are shown in Fig.~\ref{fig:stat_qsch} for quiet sun and coronal holes and in Fig.~\ref{fig:stat_ar} for active regions. In Fig.~\ref{fig:spic_density} we show distribution diagrams that compare all the regions. Table~\ref{tab:spic_stat} shows the mean and standard deviation for these nine quantities in the different regions.

\begin{deluxetable}{lrrrr}
\tablecaption{Mean spicule properties.\label{tab:spic_stat}}
\tablehead{
     & \colhead{Quiet sun} &  \colhead{Coronal hole} &  \multicolumn{2}{c}{Active region}  \\
 \colhead{Quantity} &  &  &  \colhead{(linear)} &  \colhead{(parabola)} }
\startdata
Lifetime (s)              & $108$ / $49$   & $83$   / $35$   & $165$  / $65$    & $262$   / $80$  \\
$v_{\mathrm{up}}$ ($\kms$) &   $61$ / $23$   & $71$   / $29$   & $39$   / $23$    & $30$   / $9$    \\
$l_{\mathrm{max}}$ (Mm)    & $5.48$ / $1.79$ & $6.59$ / $2.05$ & $7.75$ / $2.26$  & $6.87$ / $1.28$ \\
$H_{\mathrm{max}}$ (Mm)    & $5.48$ / $1.70$ & $6.72$ / $2.00$ & $6.91$ / $1.53$  & $6.02$ / $1.21$ \\
$\overline{H}_{s}$ (Mm)   & $2.48$ / $0.37$ & $3.09$ / $0.49$ & $2.29$ / $0.60$  & $1.91$ / $0.38$ \\
$w$ (km)                 &  $304$ / $78$    &  $318$ / $99$   &  $319$ / $95$   &  $348$ / $81$  \\
$d_{\mathrm{t}}$ (km)      & $245$  / $211$  & $342$  / $257$  &  $463$ / $402$   & $283$  / $218$  \\
$v_{\mathrm{t}}$ ($\kms$)  & $16$   / $11$   &  $20$  / $12$   & $18$   / $12$    & $14$   / $11$   \\
abs$(\theta)$ (deg)      & $12.7$ / $9.8$   & $12.7$ / $9.8$ & $18.9$ / $15.5$  & $15.3$ / $11.1$  \\
$-a$ (m\,s$^{-2}$)  & \nodata & \nodata  & \nodata & $273$/$116$ 
\enddata
\tablecomments{First value in table represents mean and second the standard deviation. $v_{\mathrm{up}}$ refers to the maximum upward velocity, $l_{\mathrm{max}}$ the maximum length, $H_{\mathrm{max}}$ the maximum height, $\overline{H}_{s}$ the mean scale height, $w$ the width, $d_{\mathrm{t}}$ the transverse displacement, $v_{\mathrm{t}}$ the maximum transverse velocity, abs$(\theta)$ the absolute value of the inclination, and $-a$ the deceleration.}
\end{deluxetable}

Spicules in the quiet sun have typical lifetimes between $50-150$~s, and slightly less in coronal holes. In both cases the distributions peak around 80~s and are noticeably asymmetric, with a sharp rise. This sharp rise could stem from selection effects in the detection of spicules, even though the minimum lifetimes are generally larger than our detection limit. In active regions spicules have much longer lifetimes. The linear spicules have a similar asymmetric lifetime distribution, but their average lifetime is longer and the distribution extends to more than 300~s. The parabola spicules in active regions have longer lifetimes, typically between $150-400$~s, and their distribution is broader and almost symmetric. 

For vertical velocity, it is in coronal holes that one finds the fastest spicules. They typically have a maximum velocity between $40-100\;\kms$, with a very small amount with velocities greater than 150~$\kms$. Quiet sun spicules have a similar maximum velocity distribution, but with a lower average and a faster drop-off after 100~$\kms$. Active regions spicules have much lower velocities. Particularly parabola spicules, whose velocities are typically between $15-40\;\kms$. The velocity of linear spicules behaves in a similar way, but peaks at a higher velocity and has a very small tail of velocities larger than $50\;\kms$.

We find similar shapes for maximum length distributions of different regions, typically centered around 7~Mm. In the quiet sun the lengths are smaller, and for coronal holes the distribution has a larger proportion of longer spicules ($> 8$~Mm). In active regions the distributions are similar to the quiet sun and coronal holes. However, when one looks at the distributions of maximum height, coronal hole spicules stand out as being the highest. This is because spicules in coronal holes tend to be less inclined, in particular compared to active region spicules. Additionally, there is a selection effect against short spicules in active regions. As mentioned in Sect.~\ref{sec:selerr}, in active regions our algorithm will preferentially detect spicules that at some point in their lives rise above the crowded `forest of spicules'. This will skew the maximum length distribution towards higher values. In some (infrequent) cases the maximum height can be larger than the maximum length. This happens when a spicule has no visible connection to the limb and its length is measured from a point above the limb. Most of these cases are found in coronal holes.

The narrowest spicule widths go down to almost the telescope's resolution limit. The extreme values measured were $138$ and $709\;$km, both in coronal hole spicules. However, most spicule diameters are found in the range of $200-450\;$km. We find very similar distributions for quiet sun and coronal hole spicules. Narrower spicules ($w<200\;$km) are less abundant in active regions. Parabola spicules tend to have the largest mean widths (by about $35\;$km compared with quiet sun and coronal holes), which is visible in their width distribution as a systematic shift towards larger values.

The distributions for transverse displacement and maximum transverse speed have similar shapes for the different solar regions. They resemble a log-normal distribution, with a peak slightly above zero. The distributions peak at a transverse displacement of about 150~km and a maximum transverse speed of about 10~$\kms$. Coronal holes stand out as having larger transverse speeds and displacements, but active region linear spicules have a greater proportion of large transverse displacements ($> 0.8\;$Mm), which gives them the largest mean. Despite having much shorter lifetimes, coronal hole spicules have larger transverse displacements than active region parabola spicules because their transverse velocities are larger. Parabola spicules tend to have the lowest maximum transverse velocities.

\subsection{Scale Height}

\paragraph{Global Scale Height}

\begin{figure}
\begin{center} 
\includegraphics[scale=0.95]{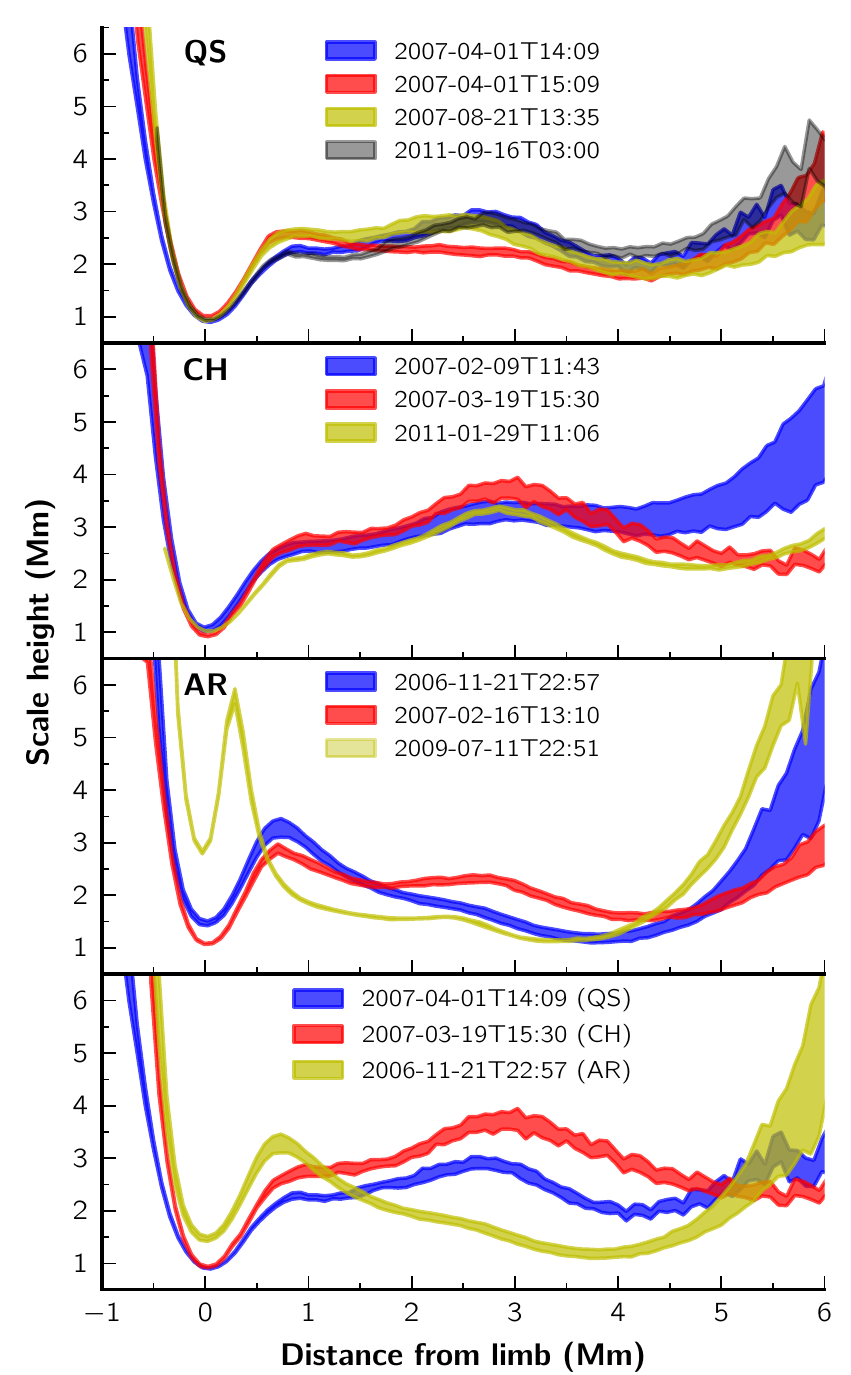}
\end{center}
\caption{Global scale height as a function of distance from limb. For each observation, bands represent the range of temporal variations within $3\sigma$ from the mean profile. Panels correspond to, from top to bottom: quiet sun (QS), coronal hole (CH), active region (AR), and a comparison between the three regions (for clarity, only one representative result from each region is shown). \label{fig:gscaleh}}
\end{figure}

Fig.~\ref{fig:gscaleh} shows the results for the global scale heights, calculated as a function of radial distance from the limb. From each set of observations, $H_s(r)$ was calculated by averaging the images in $\approx 1$~min windows, and the bands in Fig.~\ref{fig:gscaleh} represent the $3\sigma$ bounds of the temporal variations over the observing period. As the distance from the limb increases, so does the variation in these bands. This is because the S/N ratio gets progressively worse, until it becomes dominated by noise. It is not clear if the rise in $H_s(r)$ for $r \gtrsim 5$~Mm is real or an artifact of the low signal. Data sets with higher cadence have less variation in these bands because of better S/N in the $\approx 1$~min windows. In the cases of coronal holes and quiet sun, different observations show very similar scale height profiles, but in active regions there is a considerable variation between observations.

From the solar disk, scale heights fall abruptly until a minimum value of about 1~Mm at the Ca \textsc{ii} filtergram limb. Then they rise sharply again up to $2-3$~Mm at around $r\approx 0.75$~Mm, and after that their behavior varies according to the region. The scale height profiles in coronal holes and quiet sun are very similar in shape, but coronal holes tend to have scale heights larger by about $0.5$~Mm. The results for active regions illustrate how different the conditions can be for different active regions. The 2009-07-11 observations are notoriously different from any other, and even the minimum at the limb is much larger. This is most likely due to the presence of bright plage very close to the limb, which decreases $\mathrm{d}I/\mathrm{d}r$ and increases the scale height. In contrast with the quiet sun and coronal hole observations, after reaching a local scale height maximum at $r\approx 0.75$~Mm, the scale height in active regions decreases until $r\approx 3.5-4.5$~Mm. The scale heights in active regions seem to be affected by the kind of spicules detected (parabola or linear). For the 2007-02-16 observations we detect about 68\% of linear spicules, and the scale height profile is more in between the quiet sun/coronal hole and the active region sets that have mostly parabola spicules. For example, it still shows a decrease in scale height after $r\approx 0.75$~Mm, but also displays the characteristic bump at $r\approx 2.75$~Mm that is visible in most of the quiet sun and coronal hole observations.

\paragraph{Spicule Scale Height}

The spicule scale height is a more localized measure of the scale height. Calculated by a fit to the $\ln I(r)$ relation along the spicule, it is a measure of the mean scale height along the length of a spicule, up to 6~Mm. The results, in Figs.~\ref{fig:stat_qsch}, \ref{fig:stat_ar}, and \ref{fig:spic_density}, again show a significant difference between different regions. Coronal holes stand out by having the largest scale heights, on average about 0.5~Mm higher than quiet sun, in agreement with the results for the global scale heights. The active region parabola spicules have the lowest scale heights, with the distribution peaking at around 2~Mm. This is also in agreement with the results for the global scale heights. The active region linear spicules show a broader distribution. The bulk of the measurements are about 0.3~Mm larger than parabola spicules, but there is also a tail of large scale heights extending to about 4~Mm.

\subsection{Active Region Spicules\label{sec:ar_spic}}

Spicules in active regions show a broad range of properties, here categorized broadly into linear and parabola spicules based on their vertical behavior (only up or up and down, respectively). The wealth of different phenomena and processes taking place in active regions affects spicules and associated structures like macrospicules, prominences and ejections. Even though care was taken to isolate spicules from other similar structures, it is difficult to completely rule out contamination in our analysis. Adding the fact that different active region data sets vary considerably in spicule morphology and behavior, one should be careful in generalizing the behavior of spicules in any active region. 

\begin{figure}
\begin{center}
\includegraphics[scale=0.95]{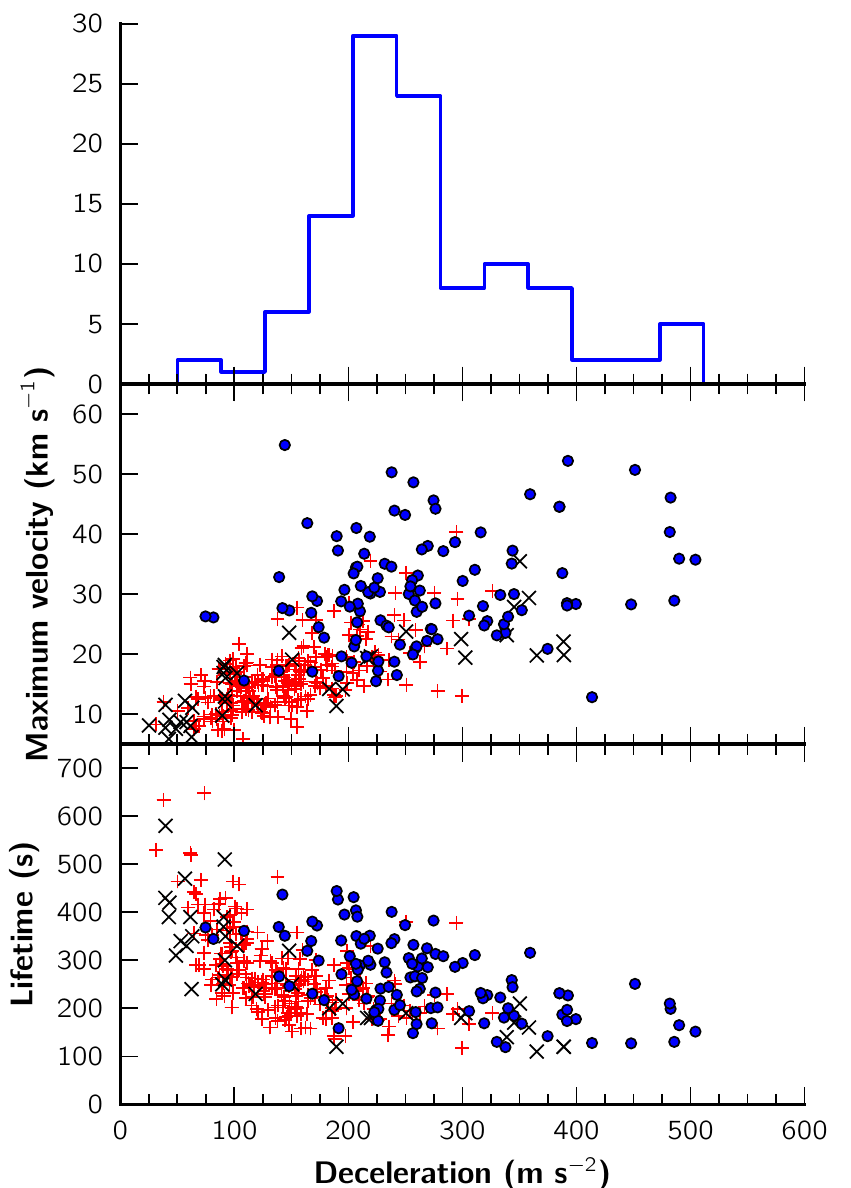}
\end{center}
\caption{Deceleration from active region parabola spicules. \emph{Top:} histogram; \emph{middle and bottom:} deceleration as a function of maximum velocity and lifetime (blue circles), compared with results for quiet sun mottles (black x markers) and active region fibrils (red crosses).\label{fig:decel}}
\end{figure}

For the parabola spicules we computed their deceleration from space-time diagrams. In Fig.~\ref{fig:decel} we show the deceleration histogram for these spicules, along with plots of the spicule maximum velocity and lifetime. For comparison these also include data for quiet sun mottles and active region fibrils from the disk observations of \citet{DePontieu:2007DF}.

Given the disparity in properties between active region spicules (even the linear kind) and spicules in quiet sun and coronal holes, we looked carefully into active region data sets to try and identify spicules that behaved more closely to those found outside active regions. We found what could possibly be a missing link. In active region observations, one often finds faint and transient objects that resemble spicules. They show similar thickness but are very faint and much shorter lived than typical spicules. At the very limb they are very faint and almost invisible in one static frame, but can be seen flickering in movies. However, they are more visible just inside the solar disk, perhaps because the superposition from limb spicules is much smaller. Being short lived and very faint, these spicules are notoriously hard to measure. 

\begin{figure}
\begin{center}
\includegraphics[scale=0.93]{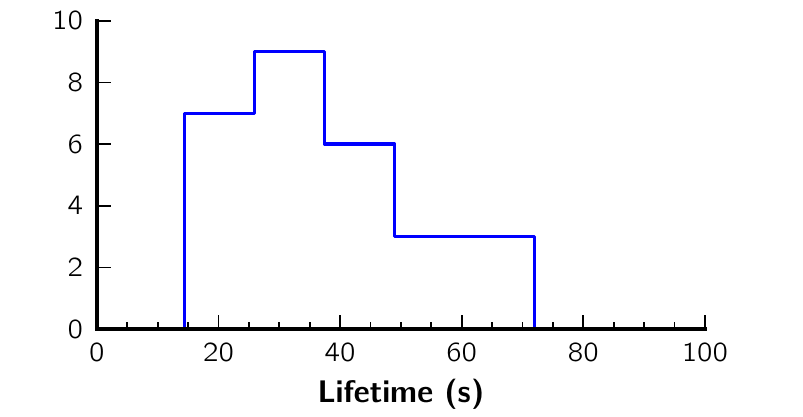}\\
\includegraphics[scale=0.93]{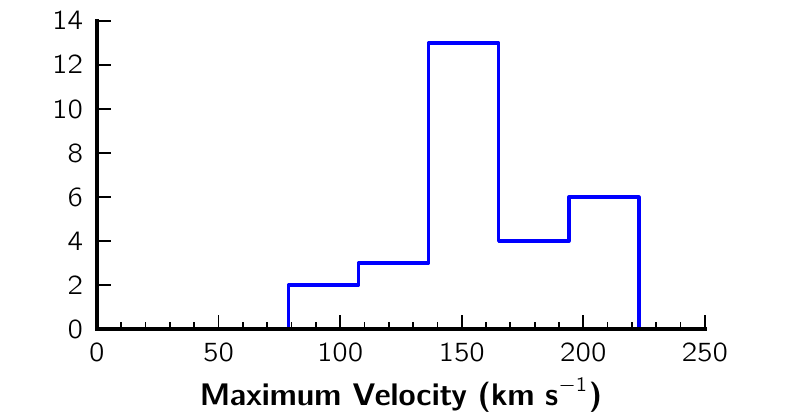}\\
\includegraphics[scale=0.93]{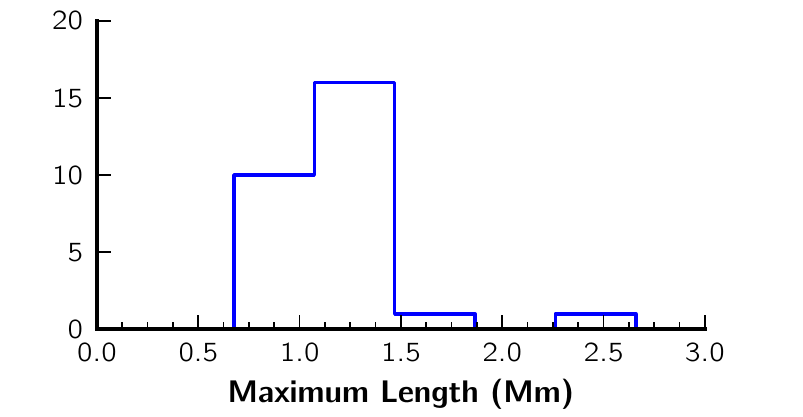}\\
\end{center}
\caption{Lifetime, maximum velocity and length histograms for the faint, quick spicules seen near active regions on the solar disk (see text).\label{fig:small_spic_hist}}
\end{figure}

To measure these faint and short-lived spicules in active regions we use additional data from observations on \mbox{2007-05-25}. In the other used active region sets these spicules are either not visible or the observational cadence is too low to track them (\emph{e.g.} the 2006-11-21 set). Because our automatic procedure struggled with these faint objects, manual detection was used on a frame-by-frame basis. Only three properties were measured: lifetime, maximum length, and maximum velocity. From this data set 28 spicules were measured. The mean $\pm$ standard deviation values were a lifetime of $37\pm16$~s, a maximum length of $1.16\pm0.35$~Mm, and a maximum velocity of $160\pm36$~$\kms$. Distributions for these quantities are given in Fig.~\ref{fig:small_spic_hist}. Systematic errors for these quantities, in particular maximum velocity and length, are most likely larger than the statistical error. This is because by being very faint, the starting and ending points of these spicules are ill-defined. Thus, the mean values should be taken with caution. We present the measurements for these spicules to illustrate that such short-lived fast objects can exist in active regions, not as a precise measurement.

It could be argued that these fast short-lived spicules come from a nearby coronal hole or quiet sun region, where spicules are more dynamic. Looking at \emph{Hinode}/XRT observations around the \mbox{2007-05-25} set we find no nearby coronal hole (the closest is more than $200\arcsec$ away). The fact that these faint objects are also seen flickering at the very limb and are present in other active region observations suggests that they are unlikely to come from a nearby coronal hole or quiet sun region.

\subsection{Qualitative Findings\label{sec:qual_find}}

Besides the quantitative properties of spicules detailed so far, there are many aspects about spicules that are difficult to quantify. Nevertheless, it is important to mention some of these findings and we detail them below. Most of these are related to what is perceived as a typical spicule, and how spicules form and evolve from an observational perspective.

\paragraph{Rise and Fall}
So far we have explicitly or implicitly assumed an idealized behavior of a spicule's life. This view entails a spicule as a linear structure that starts as a point near the limb, and then rises to a maximum extent from which it can either disappear (linear spicule) or decrease in length until it is very small (parabola spicule). While one can see examples of this behavior in the data, one cannot simply follow the evolution of most spicules. Whether because of spicule superposition, transverse motion, or low cadence, for the majority of spicules the `rise' phase is difficult to see on an individual level. Only using space-time diagrams can one identify the typical behavior more clearly. Moreover, linear spicules in particular (including many that we tracked individually) do not always disappear as soon as they reach their highest point. In many cases these spicules have a quick rise phase and then just stay at around their maximum length for several frames. This does not happen with parabola spicules, which are more regular and always fall after their maximum point.

By fitting lines to the x-t diagrams of linear spicules we assume that they are not accelerated. While a line is a good fit for most of the x-t diagrams, acceleration cannot be ruled out completely because of all the uncertainties in building the x-t diagrams and measuring velocities. A few cases hint towards an increase in the spicule velocity as it evolves, but this behavior is not generalized and it is difficult to quantify.

\paragraph{Group Behavior}
For a given field of view, despite being ubiquitous, spicules tend to originate from the same points. One can commonly observe a spicule rising from almost exactly the same footpoint as another spicule, after the other spicule disappeared or even during its lifetime. In most of these cases the spicules starting from the same point have the same inclination, although other properties can vary. This helps explain why, despite having measured hundreds of spicules, the inclination distributions are not centered at zero and for active regions are not very symmetric. The automatically detected spicules simply tended to come from similar footpoints, and the distribution of inclination was more limited than a random sampling of the field.

Besides spicules starting from similar foot points, another kind of group behavior is prevalent in active region parabola spicules. At first glance these kind of spicules may seem to be thicker, but in many cases what one sees is a cluster of spicules all rising up at similar times. In some clusters one can discern five spicules all rising at once, in other cases only two. Some examples can be seen in Fig.~\ref{fig:spic_evo}. In this figure, both spicules D and E show several spicules together. Spicule E is relatively wide and seems to have two components that move at almost the same time. Spicule D also seems to be composed of two spicules, but one of the components starts at a slightly later instant and is shorter when the first component hits the maximum length (\emph{i.e.} at around $t=240$~s). In its corresponding space-time diagram, one can see a faint evidence of two overlapping parabolas, with one having started perhaps 50~s later. The case of spicule C is more intriguing. From the space-time diagram this spicule seems like a linear spicule, even though it is hard to see it disappearing across its length at the same time. In the images of spicule C one can see a multi-component spicule (in particular after $t=280$~s).
This kind of group behavior is seldom seen in linear spicules. However, some linear spicules have two or more strands (see Spicule A in Fig.~\ref{fig:spic_evo}). It is not clear if this effect is one spicule splitting during the course of its evolution, or if more than one spicule evolve closely spaced.

Spicule B in Fig.~\ref{fig:spic_evo} shows a rare example of a new brightening starting along an existing coronal hole spicule. Starting around $t=15$~s, a bright region comes from the limb and rises with the same inclination as another spicule in the same place.  It is unclear if these are the same spicule, two spicules starting from the same point, or two spicules overlapped in this limb region. The brightening is much more intense than the existing spicule, and moves up with a velocity of about 114~$\kms$, reaching a length of about $5.7$~Mm in 56~s, all consistent with coronal hole spicules. Another characteristic of spicule B is that the bottom part of the spicule, or at least the brightest part, also moves up. This can also be seen very clearly in Spicule E, whose footpoint moves up with the spicule, having started on disk and moving up away from the limb by the end of the spicule's life.

\paragraph{Disappearance of Linear Spicules}
Another important issue with linear spicules is how exactly they disappear. We found many examples of these spicules completely disappearing over their whole length at almost the same time. In most cases this happens very quickly, in only a few images. Typically, these spicules become a little fainter and thicker just before disappearing, seemingly dissipating in the surrounding region. This cannot be observed for all spicules, in some cases they seem to `go behind' another spicule and are not seen anymore. 

\section{Discussion}
\label{sec:discussion}

\subsection{Are There Two Kinds of Spicules?}

A pressing question is if there is more than one kind of spicules. While \citet{Beckers:1968} had already considered the division of spicules into `type I' and `type II', his distinction was based on equivalent widths of spectral lines and of a totally different nature to the more recent discussion\footnote{\citet{Avery:1970} showed that the Beckers `type I' and `type II' spicules are the same features with different rotational velocities.}. \citet{DePontieu:2007} identified two fundamentally different kinds of spicules: type I (characterized by life times of $3-7$ minutes, velocities between $15-40$ $\kms$, and a parabolic up and down motion) and type II (characterized by life times of $10-150$~s, velocities between $50-150$~$\kms$, and a motion only up). During this analysis we aimed to be neutral on this question, only making a behavioral distinction between the spicules that go only up (linear) and the spicules that go up and down (parabola). However, looking at the distributions of spicule properties it is clear that the parabola and linear are two very different populations of spicules. Parabola spicules are much longer lived, have considerably lower velocities and scale heights, and exhibit group behavior very different from linear spicules. We confirm the findings of \citet{DePontieu:2007} in that the parabola spicules correspond to type I spicules and the linear spicules correspond to type II spicules. Moreover, we find that type II spicules are by far the most common type of spicule, and that type I spicules are rarely found outside active regions. 

Recently, \citetalias{Zhang:2012} have analyzed the same observations as \citet{DePontieu:2007} and draw opposite conclusions: that there are no type II spicules, and that most spicules are long-lived. In Sect.~\ref{sec:compare}, we compare and discuss both of our results, along with results from several other studies.

The discussion about the existence of type II spicules is important because they have a much larger potential to transfer energy and mass from the photosphere to the chromosphere and corona. While spicules have long been known to carry a significant mass flux to corona heights \citep{Beckers:1968} and models have been proposed in which they play important role in the energy balance of the corona \citep{PneumanKopp:1978}, their part in coronal heating has been dismissed due to a lack of a coronal counterpart \citep{Withbroe:1983}. However, \citet{DePontieu:2009} suggested that type II spicules indeed have a coronal counterpart (visible in EUV spectra), and can play an important role in driving the solar wind and supplying the corona with mass and energy. The lack of an observed downfall of type II spicules is compatible with the view that they are being heated out of the \caii\ H passband, and are the initial cool phase of spicules that are heated into transition region and coronal temperatures. Using observations from the Atmospheric Imaging Assembly (AIA) onboard the Solar Dynamics Observatory (SDO), \citet{DePontieu:2011} report a coronal counterpart of type II spicules directly in imaging, showing how some spicules evolve from \caii\ H images into higher temperature passbands in AIA, confirming their potential for heating the corona.

\subsection{Regional Differences}

Based on our findings, there are clear differences between spicules in quiet sun, coronal holes, and active regions. The results suggest that these differences are mediated by different magnetic field configurations. The differences between quiet sun and coronal hole spicules are small but measurable. In coronal holes, spicules are longer, slightly faster, shorter lived, and slightly less inclined. In coronal holes, spicules also exhibit larger transverse speeds and displacements. Despite these differences, quiet Sun and coronal hole spicules show remarkably similar dynamical behavior, with linear spicules dominant in both. From these general similarities, it is plausible that quiet sun and coronal hole spicules are the same phenomena, just exposed to different magnetic field conditions. For example, the smaller inclination in coronal holes can be expected given the open and generally more radial field configuration in coronal holes. The stronger transverse motions in coronal hole spicules could be caused by the lower rigidity of the field because of the decreased average field strength in coronal holes. The slight differences in spicule properties in the different magnetic field configurations of quiet Sun and coronal hole can thus shed light and provide constraints for models of spicule formation.

A telling example of how the properties and dynamic behavior of spicules critically depend on the magnetic field are the observations on 2007-02-09. In these equatorial limb observations, the region analyzed is squeezed in between active regions NOAA 10940 and 10941. Images from \emph{Hinode}/XRT show significant X-ray emission, and the \caii\ H filtergrams show several prominences and large ejecta above the limb, so on a first glance these data should be classified as active region. However, \citet{Liu:2011} have analyzed an impressive jet in these observations, and using a potential field source surface (PFSS) model conclude that the region is actually a localized coronal hole. This agrees very well with our findings: despite being so close to active regions, all the spicules we measured are essentially type II spicules, behaving very much like in a typical coronal hole (the global scale height profile underpins this). It shows how the local field conditions, even in such a narrow coronal hole, can completely change the spicule behavior and transition from type I spicules in active regions to type II spicules.

The classification of spicules in active regions is challenging. We measured a total of 58 linear and 112 parabola spicules in active regions, in addition to the faint and short-lived spicules near the limb. As one can see in Table~\ref{tab:spic}, the proportion of linear to parabola spicules depends on the data set. Also, aside from lifetime and transverse motion, in active regions the linear spicules behave very similarly to parabola spicules. Are these the same phenomenon? It is unclear. Others (\citealt{Anan:2010}; \citetalias{Zhang:2012}) have found spicules in active regions whose upward motion resembles a parabola but where the downfall is not observed. We did not detect such spicules in our data. In any case, the linear spicules in active regions behave much more like the parabola spicules than the type II spicules seen in the quiet sun and coronal holes. It is possible that they are indeed also type I spicules, but either dissipate before reaching the downward phase or that the absence of a downfall results from difficulties in the detection. Given the large superposition at the limb, these linear spicules could also be type II spicules from a nearby region. At the individual level it is difficult to find spicules with type II properties in active regions. Nevertheless, it seems plausible that these linear spicules could be a mix of partially-detected type I spicules and type II spicules from other regions.

The missing link between type II spicules and active regions could very well be the faint and short-lived spicules observed near the limb. These short-lived objects are notoriously difficult to measure. Our measurements provide but a rough estimate of their properties. However, they demonstrate that objects resembling spicules with high velocities ($>100\;\kms$) and short lifetimes do exist in active regions, and are possibly type II spicules. The smaller lengths found ($< 2\;$Mm) are most likely because these objects were observed on disk, where the faint tops of spicules are not visible \citep[the same effect influenced the type I spicule measurements of][]{Anan:2010}.

\begin{deluxetable*}{llcccccccc}
\tablecaption{Comparison of typical spicule properties.\label{tab:spic_others}}
\tablehead{
                      & \colhead{Region} & \colhead{Diagnostic} & \colhead{Lifetime}  & \colhead{$v_{\mathrm{up}}$}  & \colhead{$l_{\mathrm{max}}$} & \colhead{$w$}  &  \colhead{$v_{\mathrm{transv}}$}  & \colhead{abs$(\theta)$}  & \colhead{$-a$}  \\
                      &      &            &       \colhead{(s)} &         \colhead{($\kms$)} &              \colhead{(Mm)} & \colhead{(km)}      &     \colhead{($\kms$)}  &         \colhead{(deg)}  & \colhead{($m\; s^{-2}$)}}
\startdata
\citet{Beckers:1972}   & \nodata & H$\alpha$ & $250$ &            $25$ &        $6.5-9.5$ & $400-1500$ &              $<5$  &        $15-20$\tablenotemark{a} & \nodata\\
\citet{DePontieu:2007} & CH &  \caii\ H  &  $20-90$ &         $50-150$ &        $1.0-7.0$ &  $120-720$ &$0-30$\tablenotemark{b}&        \nodata & \nodata  \\
\citet{DePontieu:2007} & AR &  \caii\ H  &$180-420$ &          \nodata &          \nodata &  $120-720$ &    \nodata &        \nodata & $50-400$ \\
\citet{Pasachoff:2009} & QS & H$\alpha$  &$288-564$ &           $9-45$ &        $5.2-9.2$ &  $460-860$ &    \nodata &     $8.4-38.4$ & \nodata  \\
\citet{Anan:2010}      & AR &  \caii\ H  &    $179$ &             $34$ &            $1.3$ &  \nodata   &    \nodata &        \nodata & $510$    \\
\citet{Zhang:2012}     & CH &  \caii\ H  &    $112$ &             $41$ &            $9.6$ & $\approx 200$ & \nodata &        \nodata & $1040$   \\
\citet{Zhang:2012}     & AR\tablenotemark{c} & \caii\ H   & $176$ & $16$ &     $5.0$ & $\approx 200$ & \nodata &        \nodata & $130$    \\
This work              & QS &  \caii\ H  &$57-165$  &          $37-83$ &        $3.7-7.3$ &  $226-382$ &     $5-27$ &     $3.1-22.6$ & \nodata  \\
This work              & CH &  \caii\ H  &$45-125$  &         $42-100$ &        $4.5-8.6$ &  $219-417$ &     $8-32$ &     $2.9-22.4$ & \nodata  \\
This work              & AR &  \caii\ H  &$182-343$ &          $22-39$ &        $5.3-8.5$ &  $267-428$ &     $3-25$ &     $4.2-26.4$ & $182-343$
\enddata
\tablenotetext{a}{\citet{Beckers:1968}}
\tablenotetext{b}{\citet{DePontieu:2007sci}}
\tablenotetext{c}{Classified by \citet{Zhang:2012} as QS.}
\tablecomments{Explanation of symbols as in Table~\ref{tab:spic_stat}. QS, CH, AR refer to quiet sun, coronal hole and active region, respectively. For active regions showing only results for parabola spicules. For our results the range of values indicates the mean $\pm$ standard deviation. For other authors, range is as indicated by them. When only a single value is used, it denotes the mean.}
\end{deluxetable*}

\subsection{Comparison with Previous Studies\label{sec:compare}}

There is a wealth of previous observations of spicules and studies spanning more than 40 years. The pre-\emph{Hinode} work has been covered in many reviews \citep[\emph{e.g.}][]{Beckers:1968,Beckers:1972,Suematsu:1998}. These older observations lacked the spatial and temporal resolution of \emph{Hinode}/SOT and thus are not as reliable. A direct comparison with our results is made difficult because many of the works mentioned by \citet{Beckers:1968} and \citet{Beckers:1972} used observations in H$\alpha$ and a few other lines -- few observations were in the Ca\,\textsc{ii} H line we have used. Spicules have different lengths and different velocities in different lines, because the source functions have different depth dependencies \citep[see][who study the properties of spicules observed on disk in H$\alpha$ and Ca\,\textsc{ii} 854.2 nm]{Sekse:2012}. Nevertheless, it is relevant to compare our results with the `classical' views on spicules. 

In Table~\ref{tab:spic_others} we compare our results for the typical spicule properties with \citet{Beckers:1968,Beckers:1972}, \citet{DePontieu:2007}, \citet{Pasachoff:2009}, \citet{Anan:2010}, and \citetalias{Zhang:2012}. \citet{Pasachoff:2009} used observations in H$\alpha$, while all the other recent studies used \caii\ H from \emph{Hinode}/SOT. From Table~\ref{tab:spic_others} one can see that there is a considerable amount of variation in the literature. Some of these differences are discussed below.

The largest differences between our work and \citet{Beckers:1968,Beckers:1972} are on the spicule lifetimes, velocities (both ascending and transverse), and widths. For the lengths and inclinations our results agree reasonably. The Beckers compilations include several observations, and represent the average sun (results for particular regions are not given). When comparing with our results, the lifetimes and velocities are comparable only with the active region spicules, which were clearly not the target of those reviews.  Given the seeing-limited and low temporal and spatial resolution observations that were used, it is perhaps not surprising that the results are different. However, more recently \citet{Pasachoff:2009}, using the state-of-the-art Swedish Solar Telescope \citep[SST,][]{Scharmer:2003}, have also obtained very long lifetimes (in fact, some of the longest spicule lifetimes published for quiet sun), and relatively low velocities. Could this difference be attributed to these being H$\alpha$ observations, as opposed to \caii\ H? There are some differences between H$\alpha$ and \caii\ H spicules. \citet{Beckers:1972} notes that spicules in \caii\ H and K seem to be wider than in H$\alpha$, but admits that this may be due to different spatial resolution and seeing from early observations. \citet{Pasachoff:2009} find the opposite by measuring spicules that are almost twice as wide in H$\alpha$ than in our \caii\ H results. On the other hand, the heights measured by \citet{Pasachoff:2009} are consistent with our measurements, and the simultaneous observations in H$\alpha$ and \caii\ H of \citet{McIntosh:2008} show similar heights and dynamic behaviors. It seems unlikely that the observation of spicules in H$\alpha$ or \caii\ H can explain the differences between the classical spicules and our results. Because of the importance of this differences, we have studied this issue in a separate paper. In \citet{Pereira:2012spiclett} we degrade \emph{Hinode}/SOT \caii\ H filtergrams and find that low temporal and spatial resolutions (whether intrinsic or seeing-induced), together with the transverse motions of spicules can trick the observer (or even the automated detection) into seeing longer lived spicules that in reality are another nearby spicule. This analysis also shows that with long cadence and lower resolution observations one should expect quiet sun spicule lifetimes around $250-450$~s and velocities around 20-30~$\kms$, in much better agreement with the classical studies. The observations of \citet{Pasachoff:2009}, despite using more modern instrumentation, are still affected by seeing and have a cadence of about 50~s, which will be affected by the sampling effects. According to our results, cadences of this order are clearly insufficient for a precise measurement of quiet sun spicules. \citet{Anan:2010} also use a low cadence of 45~s, but because the target is active region plage, where spicules are longer lived, its effects are probably not as serious. 

Our results compare favorably with \citet{DePontieu:2007} and with the transverse displacements and velocities given by \citet{DePontieu:2007sci}. This agreement is pleasing but not surprising given that their observations are a subset of ours.

\citet{Anan:2010} have observed spicules near the limb but on the solar disk. Their results are peculiar, in particular the very small lengths. With mean lengths of 1.3~Mm, they report the shortest spicules from all these studies. However, this is not so surprising because these spicules were observed on disk. With a nearly exponential intensity profile along their lengths, the tops of spicules become very faint. At the limb they are just visible near the noise level, but on disk filtergrams the signal will be dominated by upper photospheric sources and only the bottom part of the spicules is visible. This is the most likely reason for this large discrepancy in spicule length, and likely also the lifetime (ours are longer). One should thus exercise caution when comparing our work with \citet{Anan:2010}, because we are measuring different parts of the spicules. Nevertheless, their velocities seem consistent with our active region velocities, even though our accelerations are considerably lower and our lifetimes longer.

\begin{figure}
\begin{center}
\includegraphics[scale=0.95]{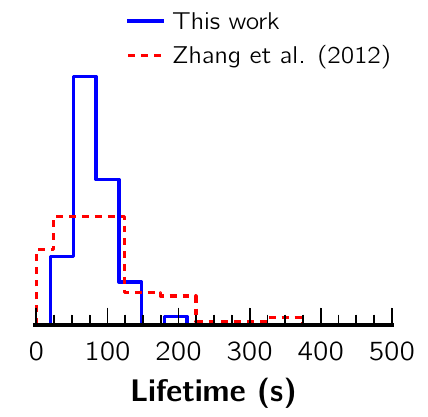}\includegraphics[scale=0.95]{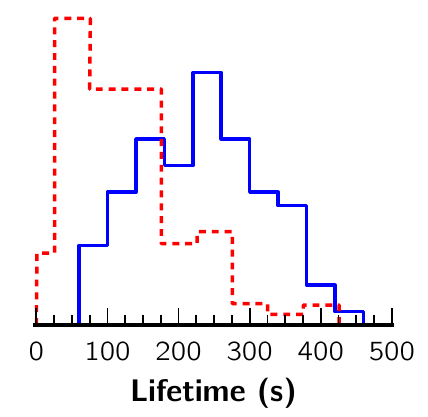}\\
\includegraphics[scale=0.95]{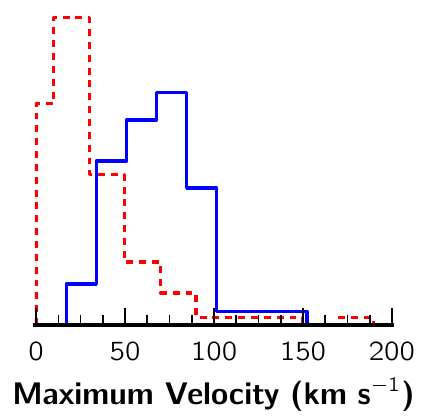}\includegraphics[scale=0.95]{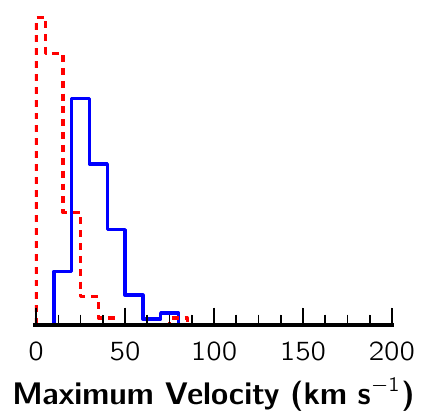}\\
\includegraphics[scale=0.95]{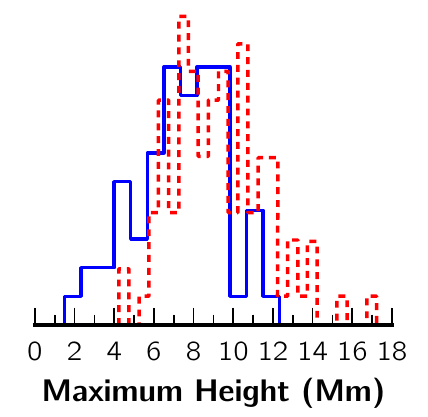}\includegraphics[scale=0.95]{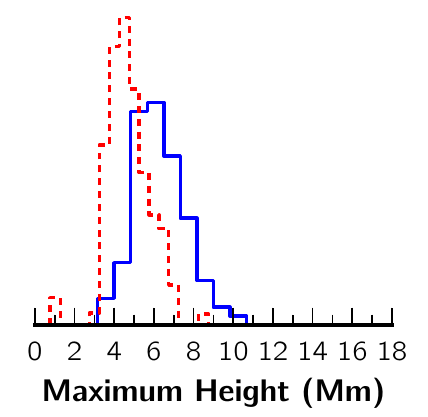}\\
\end{center}
\caption{Normalized histograms for our results and those of \citet{Zhang:2012}, for the coronal hole data set of 2007-03-19 (\emph{left column}) and the active region data set of 2006-11-21 (\emph{right column}).\label{fig:zhang}}
\end{figure}

A comparison with \citetalias{Zhang:2012} is very interesting because we have similar aims and use almost the same data. \citetalias{Zhang:2012} have analyzed a subset of the observations we used, namely the active region of 2006-11-21 and the coronal hole of 2007-03-19. As for the latter, we have used the observations starting at 15:30 UT, the same observations as \citet{DePontieu:2007}. \citetalias{Zhang:2012} used observations of the same field but taken four hours earlier, starting at 11:29 UT. These two observations have the same pointing and observational parameters. As seen in movies they are almost identical in spicule behavior; for our purposes we will therefore assume that they are the same. 

In Fig.~\ref{fig:zhang} we compare the results of \citetalias{Zhang:2012} with our results for the same observations. \citetalias{Zhang:2012} measure spicule heights instead of lengths, and their Vy is the maximum velocity in the vertical direction and not along the spicule. Because spicules are inclined, their Vy will always be lower than the true velocity. In terms of maximum heights, our results are not too dissimilar from \citetalias{Zhang:2012}. We have smaller heights for the coronal hole set and longer heights for the active region set. Given that the height is somewhat subjective (the top end of spicules is very faint, ill-defined, and likely dependent on the radial filtering and image processing, and the limb position not always clear), such differences are perhaps to be expected. As for lifetime, our results agree reasonably for the coronal hole, but we get substantially larger lifetimes in the active region. The largest disagreements are found in the maximum velocities, where we systematically find much larger values, about twice as large. This difference cannot be explained by the vertical velocity projection alone.  

However, the most striking disagreement we have with \citetalias{Zhang:2012} is on the spicule behavior in the coronal hole. For the active region observations, \citetalias{Zhang:2012} find 71 parabola and 9 linear spicules, whereas we find 79 parabola and 19 linear, which is consistent. But for the coronal hole observations we find only linear spicules (60), while \citetalias{Zhang:2012} find 64 parabola spicules and 24 linear spicules. They claim a 63\% fraction of parabola spicules, which we could not find. 

Even though \citetalias{Zhang:2012} used a radial filter similar to ours, perhaps a different image scaling could lead one to see apparent downward motion of some spicules. Looking at movies of these observations repeatedly, we can only find a handful of spicules going down (and certainly not such a high proportion) but they do not seem to follow parabolic paths. \citetalias{Zhang:2012} give a single example of one such parabola spicule (their Fig.~5), but to us the filtergrams are fuzzy and unconvincing.

For the parabola spicules in coronal holes \citetalias{Zhang:2012} derive a significant mean deceleration of 1.37~km\,s$^{-2}$, much larger than our mean deceleration of 273~m\,s$^{-2}$ in active regions. On the other hand, their measured deceleration of active region parabola spicules is only 130~m\,s$^{-2}$. Additionally, the mean values given by \citetalias{Zhang:2012} are inconsistent. For the active region parabola spicules their mean values are $t=176$~s, $v_{\mathrm{up}}=16.9\:\kms$, and $a=-130$~m\,s$^{-2}$. A parabola with a starting velocity of $v_{\mathrm{up}}$ and acceleration $a$ will have a lifetime of 260~s, much longer than the mean lifetime of 176~s. For our active region mean values of $t=262$~s, $v_{\mathrm{up}}=30\:\kms$, and $a=-273$~m\,s$^{-2}$, a parabola would have a lifetime of 220~s, reasonably close to our mean lifetime. For the coronal hole, \citetalias{Zhang:2012} have mean values of $H=9.6$~Mm, $v_{\mathrm{up}}=48\:\kms$, and $a=-1.37$~km\,s$^{-2}$. A parabola with this $v_{\mathrm{up}}$ and $a$ would have a lifetime of 70~s, compared to their average lifetime of 121~s. To achieve the maximum height of $H=9.6$~Mm, let us assume that the average vertical speed is the mean derivative of this parabola, $v=v_{\mathrm{up}}/2 = 24\:\kms$. Dividing 9.6~Mm by this value gives a lifetime of 400~s. Even assuming that the spicule was not a parabola but a linear spicule going up with $v_{\mathrm{up}}$, it would still take about 200~s, a very different value from their mean life time of $121$~s. For the few linear spicules that \citetalias{Zhang:2012} found in the coronal hole, mean values were $H=10.4$~Mm, $v_{\mathrm{up}}=27\:\kms$. In this case a spicule rising linearly would take $390$~s, again a very different value from their mean lifetime of $100$~s. For our coronal hole mean values, dividing $H=6.6$~Mm by $v_{\mathrm{up}}=70\:\kms$ one gets $94$~s, reasonably close to our mean lifetime of $85$~s. We find these inconsistencies in the work of \citetalias{Zhang:2012} troubling.

Furthermore, \citetalias{Zhang:2012} criticize the use of space-time diagrams to derive spicule velocities by \citet{DePontieu:2007} as a bad approximation that leads to erroneous trajectories of spicules. They seem to ignore the fact that the analysis of \citet{DePontieu:2007} specifically excludes spicules with transverse motions (something that is clearly stated in the paper) and thus avoids any issues with space-time diagrams. Our visual inspection of Ca II H movies does not show evidence for the dominance of up and downward motion that they claim to see in coronal holes. \citetalias{Zhang:2012} further claim that a majority of type II spicules does not exist when one traces the spicule properly. As our work shows, by self-consistently tracking the spicule and its transverse motion it is possible to find many examples of type II spicules, and they are by far the most common type.

\subsection{Disk Counterparts and Driving Mechanisms}

The discussion so far pertains spicules seen at the limb. To understand the origin of spicules it is important to identify how they look on the solar disk. Different phenomena have been observed on disk, and provide plausible explanations to type I and type II spicules.

\paragraph{Type I spicules} Disk observations of active region dynamic fibrils (DFs) show jet-like features with parabolic ascending and descending motions. The same is seen for mottles, which are believed to be the quiet sun counterparts of DFs \citep[see][]{vdVoort:2007}. While mottles are longer than DFs, both have similar properties.  Using SST H$\alpha$ observations, \citet{DePontieu:2007DF} measure 257 DFs and find lifetimes between $100-500$~s, maximum lengths between $0.5-3$~Mm, maximum velocities between $10-30$~$\kms$, and decelerations between $30-300$~m\,s$^{-2}$. Using also H$\alpha$ observations, \citet{vdVoort:2007} derive similar properties for mottles, consistent with earlier results from \citet{Suematsu:1995}. Given their similar properties and behavior, mottles and DFs have been proposed to be the disk counterparts of type I spicules \citep{Tsiropoula:1994,Suematsu:1995,Christopoulou:2001,DePontieu:2007DF,vdVoort:2007,{Martinez-Sykora:2009}}. Our results support this view, with the properties of type I spicules (active region parabolas) being comparable to those of DFs and mottles. However, mottles are found on quiet sun and we find almost no type I spicules outside of active regions. This discrepancy can be reconciled by taking into account two issues. First, in contrast to active region DFs, quiet sun mottles most often occur on what appear to be inclined field lines \citep{vdVoort:2007}, likely because the magnetic field is not as structured on large-scales in quiet sun (compared to active regions). In addition, numerical modeling of type I spicules (see below) suggests that the slow mode magneto-acoustic shocks thought to be responsible for these jets are significantly weaker, and thus drive plasma to significantly lower heights, in magnetic field conditions that are representative of quiet sun or coronal holes. This is because the shock energy is diluted over a larger area as the magnetic field in quiet sun expands much more with height than in active regions (since the average coronal field in an active region is much larger). The combination of both issues plausibly leads to jets that reach significantly lower heights under quiet sun and coronal hole conditions, leading to a lack of visibility at the limb. While the limb counterparts of these shock-driven jets are not observed in quiet sun and coronal holes, they do occur in these regions, as clearly evidenced by a wealth of observational evidence on the solar disk \citep[\emph{e.g.}][]{vdVoort:2007}. 
In any case, the fact that mottles tend to be found near high magnetic flux regions \citep{vdVoort:2007} supports the idea that both mottles and dynamic fibrils arise from strong magnetic field regions.

The driving mechanism for these jets has long been debated \citep[see review by][]{Sterling:2000}. Using numerical simulations and high-resolution observations, \citet{DePontieu:2004} established that magnetoacoustic shock waves (from photospheric p-mode leakage) are an effective driving mechanism of type I spicules. This view has been reinforced by both simulations and observations \citep{Hansteen:2006,Heggland:2007,DePontieu:2007DF,vdVoort:2007,Martinez-Sykora:2009,Heggland:2011}. The observed group behavior of type I spicules (when several spicules are seen rising together) is consistent with an ascending shock front, and is supported by DF observations \citep{DePontieu:2007DF} and simulations \citep{Heggland:2011}. Given the shock wave mechanism, the simulations of \citet{Heggland:2007} confirm a linear relation between the deceleration and maximum velocity of these jets, as well as inverse relation between the deceleration and lifetime. Our results for these relations, shown in Fig.~\ref{fig:decel}, are in good agreement with the theoretical predictions and observations of mottles and DFs from \citet{DePontieu:2007DF}. A similar relation is found by \citet{Anan:2010}, although their measurements extend to much higher decelerations and velocities. Compared to the mottles and DFs, our larger velocities are likely a result of the projection effects, which are more severe on the disk data. The scatter in the values, as noted by \citet{Heggland:2007}, is a result of different excitation periods and different projection angles. These results further strengthen the view that DFs and mottles are the disk counterparts of type I spicules, and that they are driven by magnetoacoustic shocks.

\paragraph{Type II spicules} \citet{Langangen:2008} identify features on the disk, so called `rapid blueshifted events' (RBEs), and propose them as the disk counterpart of type II spicules. Using high-quality H$\alpha$ and \caii\ 854.2~nm observations of a coronal hole, \citet{vdVoort:2009} solidify this view, and \citet{Sekse:2012} further extend this analysis by detecting more coronal hole RBEs with a semi-automated method similar to our own for spicules. For the RBE properties, \citet{Sekse:2012} find lifetimes between $20-100$~s, lengths between $1-6$~Mm, and doppler velocities between $10-45\;\kms$. The doppler velocities differ for  H$\alpha$ and \caii\ 854.2~nm observations, and they are lower than the apparent velocities \citep[see][, who find apparent velocities for RBEs on the order of 50$\;\kms$]{vdVoort:2009}. These measurements are consistent with our results for type II spicules. We find very similar lifetimes between RBEs and type II spicules in coronal holes and quiet sun. Our lengths are larger by $\approx 3\;$Mm, which is reasonable given the projection effects and that the tops of spicules are too dim (lack opacity) to be observed on disk. The same projection effect and lack of opacity argument also holds when comparing our velocities, which are almost twice as high. By tracking the trajectories of spicules, \citet{Sekse:2012} also calculate the transverse displacement and transverse velocities. The shape of their distributions agrees very well with ours, although we obtain transverse displacements larger by about 150~km and maximum transverse speeds larger by about 10$\;\kms$. This discrepancy can be explained by taking into account the fact that the automated detection algorithm by \citet{Sekse:2012} imposes a maximum transverse velocity of order 11$\;\kms$ to avoid false positive identifications.  In summary, we find the agreement with the RBE studies encouraging and supportive of the hypothesis that they are the disk counterparts of type II spicules.

Unlike for type I spicules, the driving mechanism for type II spicules has not been clearly established. There are several theories proposed, ranging from magnetic reconnection \citep[\emph{e.g.}][and references therein]{Moore:2011} to Alfv\'en waves \citep[\emph{e.g.}][]{Matsumoto:2010}. Recently, some progress has been made in reproducing type II spicules from first-principles 3D rMHD simulations \citet{Martinez-Sykora:2011}. But further observational studies are needed (in particular, of the regions around the footpoints) to understand this complex phenomenon.

\subsection{Are Spicules Sheets?}

Recent work by \citet[hereafter \citetalias{Judge:2011}]{Judge:2011} has brought up the provocative idea that spicules are not manifestations of plasma motion but instead ``warped current sheets''. In what would amount to a paradigm shift regarding the spicule deposition of energy and mass in the corona, \citetalias{Judge:2011} claim that the warped current sheet explanation provides the simplest physical picture. According to the authors, the motivations for such explanation come from (1) the fluted sheet structure of photospheric magnetic structures and (2) the likelihood of magnetic tangential discontinuities to develop in the low plasma $\beta$ environment. 

While in our analysis there are some aspects that could lend support to the sheet hypothesis, in general we fail to see any significant observational proof of spicules as sheets. It should be said, nevertheless, that in the vague definition formulated by \citetalias{Judge:2011}, it will be very difficult to find conclusive observational evidence for or against this hypothesis, because of the unknown and invisible orientation of the sheets and of the many possible field configurations that theoretically could explain any single field of view. Yet, we find several inconsistencies between our observations and this hypothesis, and we detail them below.

First let us get the type I spicules out of the way. As \citetalias{Judge:2011} admit, the sheet explanation might not explain all spicules, and the authors focus on type II spicules. Given how they are consistent with a magnetoacoustic shock wave driver, a dynamic fibril disk counterpart, and a time-scale of photospheric p-modes, the current view on type I spicules is consistent and well established. So it seems most unlikely that these spicules are sheets, which \citetalias{Judge:2011} do not dispute. Their main point is that type II spicules \emph{could} be sheets. However, if type I spicules are not sheets and indeed jets, it throws away one of the main arguments for the very existence of sheets, which \citetalias{Judge:2011} summarize as ``how can one get straw-like structures out of the fluted sheet fields that appear to dominate the photospheric network (...)?''. More likely than not, type I spicules are jets, which disproves the premise that straw-like features cannot originate from heterogeneous structures in the photosphere.

One of the arguments for the sheet hypothesis is that the time-scales of type II spicules are inconsistent with the timescales of granular flows, which \citetalias{Judge:2011} estimate to be around $60-120\;$s. While \citetalias{Judge:2011} refer to the results of \citet{DePontieu:2007}, when looking at our distribution for lifetime of type II spicules, most have lifetimes between $50-150\;$s, consistent with these granular flow timescales. However, the timescale argument is unconvincing. This is because \citet{DePontieu:2011} indicate that \caii\ observations of type II spicules simply show the initial phase of a spicular event. During this initial phase the chromospheric plasma is rapidly heated and fades from the \caii\ passband, to appear in the hotter He\,\textsc{ii} $30.4\;$nm passband, where the spicules appear to continue for several more minutes. The timescales to compare to are then significantly longer than the \caii\ timescales, and they are in fact of order granular timescales, contrary to the claim by \citetalias{Judge:2011}. In other words, the \caii\ timescales are simply a measure of the heating timescales as a result of whatever process drives the upflows. Given the limited temperature range these chromospheric diagnostics are sensitive to, the heating timescales are shorter than the timescales of the underlying driving mechanism.

The sheet hypothesis requires highly corrugated sheets to explain the fine structure and closely packed thin spicules seen with \emph{Hinode}. If these spicules are indeed just a density effect like ``the fluted parts of curtains'' and are closely packed, then from the motion of the sheet one would expect to see spicules going up as well as spicules going down. But observationally the overwhelming majority of spicules (in particular in coronal holes) are only seen to go up. \citetalias{Judge:2011} attempt to explain this shortcoming by arguing that the warps are driven from below. If this was true, then one would expect the footpoints of spicules to move considerably, which is not seen. In fact, as noted in Section~\ref{sec:qual_find}, spicules tend to come from the same footpoints for long periods of time. Perhaps more importantly, the predominantly closed fields in quiet sun should allow for roughly equal amount of up and downward propagation, and thus apparent motion in the ``sheets''. This is directly contradicted by our observations in quiet sun spicules which overwhelmingly show upward apparent motions, and almost never downward apparent motion. 

One issue that is not at all addressed in the sheet hypothesis is how the cool plasma contained in spicules ``magically'' appears at what are essentially coronal heights. Under solar conditions, singly ionized calcium ions rapidly become ionized to double ionized calcium for temperatures above \mbox{20\,000\,K} \citep{Wedemeyer:2011}. This means that our \caii\ emission in spicules emanates from cool plasma that occurs at heights much in excess of the scale height expected for such plasma. How does this chromospheric plasma make it to coronal heights in excess of 10 Mm above the photosphere? Clearly it is not in hydrostatic equilibrium, and some force must propel the plasma to those heights. Yet the sheet hypothesis remains vague and does not provide an explanation for the perhaps most studied and puzzling aspect of spicules: how this cool, dense chromospheric plasma appears at such great heights.

Other observational aspects from the observations of spicules, such as the upward propagation of bright spots along the spicules (like the example spicule B in Fig.~\ref{fig:spic_evo}), also make an explanation with the sheet hypothesis extremely complicated. All things considered, the sheet hypothesis is an interesting new view on a much debated subject, but that ultimately fails to present a simpler picture of the formation of spicules, and is inconsistent with several observational diagnostics. It is likely and cannot be ruled out that such current sheets do form in the solar atmosphere, but that they are responsible for spicules seems unlikely.

\section{Conclusions}
\label{sec:conclusions}

Measuring limb spicules is a daunting task. Providing long time series of seeing-free data, \emph{Hinode} is arguably the best observatory for characterizing spicules. Using a semi-automated procedure we obtained a consistent and statistically-significant set of spicule measurements over quiet sun, coronal holes, and active regions.

We find strong evidence of two populations of spicules, type I and type II, in agreement with \citet{DePontieu:2007} and in contradiction of \citet{Zhang:2012}. The typical properties of type I spicules are lifetimes of $150-400\;$s, maximum ascending velocities of $15-40\;\kms$, maximum heights of $4-8\;$Mm, inclinations of $0-40\;$degrees, scale heights of $1.5-2.5\;$Mm, and maximum transverse speeds of $5-30\;\kms$. Type II spicules, on the other hand, have much shorter lifetimes of $50-150\;$s, larger ascending velocities of $30-110\;\kms$, maximum heights of $3-9\;$Mm (longer in coronal holes), inclinations of $0-30\;$degrees, scale heights of $1.8-3\;$Mm in quiet sun and $2.4-3.6\;$Mm in coronal holes, and slightly larger maximum transverse speeds of $4-40\;\kms$. Type I spicules rise and fall with a parabolic trajectory, while type II spicules rise linearly until their maximum length, and then dissipate quickly, typically over their whole length and suggesting that they are heated out of the \caii\ passband. We find that type II spicules are dominant in the quiet sun and coronal holes, with type I spicules being found mostly in active regions.

Our results differ from the classical views on spicules by \citet{Beckers:1968, Beckers:1972}, who reported spicules with long lifetimes and slow velocities -- resembling the properties that we measure for type I spicules. However, these comprehensive reviews include results from quiet sun and coronal hole observations that most likely included both type I and type II spicules. This puzzling discrepancy between our results and classical spicules is addressed in \citet{Pereira:2012spiclett}, where it is found that the lower resolution of earlier studies can affect the spicule measurements, and by degrading \emph{Hinode} data to similar conditions of these studies, we measure spicule properties that agree with classical spicules. Thus, we believe that this difference stems from the higher quality observations available today.

The observations of type I spicules are consistent with a magnetoacoustic shock wave driver \citep{DePontieu:2004,Hansteen:2006,Heggland:2007}, and that dynamic fibrils and mottles are their disk counterparts \citep{DePontieu:2007DF,vdVoort:2007}. Given the relations between the measured decelerations, velocities, and lifetimes, our results support this view.

The ubiquity of type II spicules lends credence to the hypothesis that they can supply the corona with energy and mass, even if only a small percentage of them reach coronal heights \citep{PneumanKopp:1978,DePontieu:2009,DePontieu:2011}. Furthermore, the suggestion by \citet{Langangen:2008} that RBEs are the disk-counterpart of type II spicules has been underpinned by \citet{vdVoort:2009} and \citet{Sekse:2012}. Nevertheless, more detailed observations on the origins of type II spicules and improved numerical models are necessary for one to grasp their complex interaction from the photosphere to the corona.

\acknowledgements

TMDP's research was supported by the NASA Postdoctoral Program at Ames Research Center through contract number NNH06CC03B. BDP gratefully acknowledges support by NASA grants NNX08AH45G, NNX08BA99G, and NNX11AN98G. This research was supported by the Research Council of Norway. \emph{Hinode} is a Japanese mission developed by ISAS/JAXA, with the NAOJ as domestic partner and NASA and STFC (UK) as international partners. It is operated in cooperation with ESA and NSC (Norway). The authors thank Silje Bj\o lseth and Anne Fox for help with the data reduction and Alan Title, Rob Rutten, Luc Rouppe van der Voort, Viggo Hansteen, Scott McIntosh, and Ted Tarbell for interesting discussions. We would like to thank the referee, whose comments were very helpful and improved the manuscript.

\bibliographystyle{apj}

\end{document}